\documentclass[
aps,
prl,
reprint,
groupedaddress,
]{revtex4-1}

\usepackage [utf8]{inputenc}

\usepackage{amssymb,amsmath}
\usepackage[
]{graphicx}
\usepackage{xcolor}
\usepackage{dsfont}

\usepackage[english]{babel}

\usepackage{siunitx}

\usepackage{pgfplots}

\DeclareMathOperator{\sgn}{sgn}

\usetikzlibrary{arrows,shapes,backgrounds,calc,
    positioning,
    intersections}

\newcommand{\ket}[1]{\left| #1 \right>} 

\newcommand{\I}{\mathrm{i}}

\newcommand{\unitn}{\hat{\mathbf{n}}}

\newcommand{\unitx}{\hat{\mathbf{x}}}
\newcommand{\unity}{\hat{\mathbf{y}}}
\newcommand{\unitz}{\hat{\mathbf{z}}}

\newcommand{\Bbold}{\mathbf{B}}
\newcommand{\Jbold}{\mathbf{J}}

\usepackage{layouts}
\usepackage{float}

\begin{document}

 \title{
Probing quantum criticality and symmetry breaking at the microscopic level
 }
 \author{Vasiliy Makhalov}
 \thanks{These two authors contributed equally.}
 \author{Tanish Satoor}
 \thanks{These two authors contributed equally.}
 \author{Alexandre Evrard}
 \author{Thomas Chalopin}
 \author{Raphael Lopes}
 \author{Sylvain Nascimbene}
 \email{sylvain.nascimbene@lkb.ens.fr}
 \affiliation{Laboratoire Kastler Brossel,  Coll\`ege de France, CNRS, ENS-PSL University, Sorbonne Universit\'e, 11 Place Marcelin Berthelot, 75005 Paris, France}
 \date{\today}
 \begin{abstract}
We report on an experimental study of the Lipkin-Meshkov-Glick model of  quantum spins  interacting at infinite range in a transverse magnetic field, which exhibits a ferromagnetic phase transition in the thermodynamic limit. We use Dysprosium atoms of electronic spin $J=8$, subjected to a quadratic Zeeman light shift, to simulate $2J=16$ interacting spins $1/2$. We probe the system microscopically using single magnetic sublevel resolution, giving access to the spin projection parity, which is the collective observable characterizing the underlying $\mathbb{Z}_2$ symmetry. We measure the thermodynamic properties and dynamical response of the system, and study the quantum critical behavior around the transition point. In the ferromagnetic phase, we achieve coherent tunneling between symmetry-broken states, and test the link between symmetry breaking and the appearance of a finite order parameter. 
 \end{abstract}
 
 \maketitle
 
From complex quantum materials such as cuprate superconductors to simple spin models, many-body systems close to a quantum critical point exhibit distinct properties driven by quantum fluctuations \cite{sachdev2011quantum}. Some features, such as the slowing down of relaxation times, can be probed via macroscopic observables. However, revealing specifically quantum properties, e.g. many-body quantum entanglement \cite{osterloh_scaling_2002}, remains challenging. 
The recent development of highly controlled quantum systems of mesoscopic size, such as  ion crystals \cite{blatt_quantum_2012}, ultracold gases \cite{gross_quantum_2017-1}, Rydberg atom arrays \cite{saffman_quantum_2016}, or interacting photons \cite{ma_quantum_2011-1}, allows for a microscopic characterization of collective quantum properties \cite{georgescu_quantum_2014-1}, e.g. the full density matrix \cite{ma_quantum_2011-1}, entanglement entropy \cite{islamMeasuringEntanglementEntropy2015} or non-local string order \cite{hilkerRevealingHiddenAntiferromagnetic2017}. This degree of control could be used to investigate fundamental aspects of quantum phase transitions, such as the link between the breaking of an underlying symmetry and the onset of a non-zero order parameter \cite{landau1937theory}. This connection cannot be tested in macroscopic systems as superselection rules forbid large-size quantum superpositions \cite{bartlett_reference_2007}, making spontaneous symmetry breaking unavoidable \cite{anderson1984basic}.

\begin{figure}
\includegraphics[
draft=false,scale=1.0,trim={1mm 2mm 0mm 0mm},
]{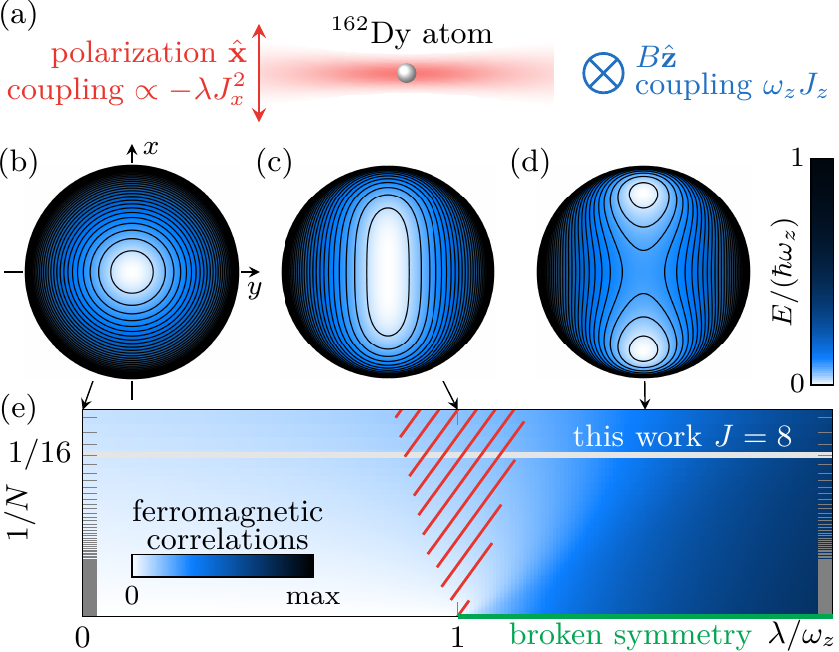}
\caption{
(a) Scheme of the experiment, based on laser-induced non-linear dynamics of the electronic spin of Dysprosium atoms (quadratic light shift $\propto-\lambda J_x^2$), in the presence of a magnetic field inducing a Zeeman coupling $\omega_zJ_z$.
(b,c,d) Classical energy landscapes calculated on the southern hemisphere of the generalized Bloch sphere, for $\lambda=0$, $\omega_z$ and $1.5\,\omega_z$ (b, c and d, respectively). 
(e) Finite-size phase diagram, with a ferromagnetic phase in the thermodynamic limit for $\lambda>\omega_z$ (green line).  For a finite $N$ the phase transition is smoothened over a quantum critical region (dashed red). 
\label{fig_scheme}}
\end{figure} 

In this Letter, we experimentally characterize at the microscopic level the Lipkin-Meshkov-Glick  model (LMGm), consisting of $N$ quantum spins  with infinite-range Ising interactions in a transverse field. This model is applicable to nuclear systems \cite{lipkin_validity_1965,cejnar_quantum_2010}, large-spin molecules \cite{gatteschi_quantum_2003}, trapped ions \cite{lanyon_universal_2011,islam_onset_2011} or two-mode Bose-Einstein condensates \cite{albiez_direct_2005,levy_.c._2007,trenkwalder_quantum_2016}.  Our study is based on the equivalence between  the electronic spin $J=8$ of  Dysprosium atoms and a set of $N=16$ spins $1/2$ symmetric upon exchange \cite{landau_quantum_1958}, with Ising interactions simulated  via a light-induced quadratic Zeeman shift \cite{smith_continuous_2004-2}. In the thermodynamic limit (TL), the LMGm exhibits a ferromagnetic phase transition (see Fig.\,\ref{fig_scheme}), characterized by spontaneous breaking of a $\mathbb{Z}_2$ symmetry -- the parity of the total $z$ spin projection.
We measure a crossover between para- and ferromagnetic behaviors, separated by a quantum critical regime where we observe non-classical behavior and a minimum of the energy gap \cite{botet_size_1982,dusuel_finite-size_2004}. A specific asset of our setup is the direct access to the quantum state parity, a collective observable hidden in macroscopic systems. We show that the $\mathbb{Z}_2$ symmetry breaking is directly related to the onset of a non-zero order parameter.

The LMGm is described by the Hamiltonian
\begin{equation}\label{eq_Lipkin}
 H=-\frac{\hbar\lambda}{4(N-1)}\sum_{1\leq i\neq j\leq N}\sigma_{ix}\sigma_{jx}+\frac{\hbar\omega_z}{2}\sum_{1\leq i\leq N}\sigma_{iz}.
\end{equation}
Here, $\hbar\sigma_{iu}/2$ denotes the projection of the spin $i$ along $u$ ($1\leq i\leq~N$), the factor $1/(N-1)$ ensures extensivity of the energy for large $N$ \cite{campa_physics_2014}, and we restrict ourselves to ferromagnetic interactions $\lambda>0$. As each spin interacts with the sum of all other spins, classical mean-field theory becomes valid in the TL \cite{botet_size_1982}. The corresponding classical energy functionals, parametrized by the mean spin orientation, are shown in Fig.\,\ref{fig_scheme}b,c,d for $\lambda=0$, $\omega_z$ and $1.5\,\omega_z$. They reveal the occurrence of a quantum phase transition between a paramagnetic phase for $\lambda<\omega_z$ and a ferromagnetic phase for $\lambda>\omega_z$, for which the system exhibits two degenerate ground states with non-zero order parameter $\langle \sigma_{1x}\rangle\neq0$. Furthermore, the $\mathbb{Z}_2$ symmetry, associated to the conservation of  parity $
P_z=\prod_{i=1}^N\sigma_{i,z}
$, is spontaneously broken at the  ferromagnetic transition.
Introducing the collective spin $\Jbold=\tfrac{1}{2}\sum_i \boldsymbol{\sigma}_i$, the Hamiltonian (\ref{eq_Lipkin}) can be recast (up to an overall energy shift
) as 
\begin{align}
H=-\frac{\hbar\lambda}{2J-1} J_x^2+\hbar\omega_z J_z.
\end{align} For ferromagnetic interactions, its lowest energy states are permutationally symmetric and their collective spin has the  maximal length $J=N/2$.

In this work, we study the non-linear dynamics of the electronic spin $J=8$ of $^{162}$Dy atoms, simulating  the collective spin of a ferromagnetic LMGm with $N=16$ spins 1/2. We use ultracold samples of $1.3(3)\cdot10^5$ atoms, initially held in an optical dipole trap at a temperature $T\simeq\SI{1.1(1)}{\micro K}$. The atomic spin is initially polarized in the ground state $\ket{-J}_z$, under a magnetic field $\Bbold=B\unitz$ with $B=\SI{114(1)}{mG}$, corresponding to a Larmor frequency $\omega_z=2\pi\times\SI{198(2)}{kHz}$. In this state, the $N$ elementary spins are anti-aligned with the magnetic field, corresponding to a paramagnetic state. We then switch off the trap  before applying an off-resonant laser beam close to the 626-nm optical resonance, linearly polarized along $x$, resulting in a quadratic Zeeman light shift $\propto J_x^2$ \cite{smith_continuous_2004-2}. After a typical evolution time $t\sim\SI{100}{\micro s}$, we switch off the light beam,  apply time-dependent magnetic fields to perform arbitrary spin rotations, before making a projection measurement along $z$. Combining rotation and projection gives us access to the spin projection probabilities, 
$\Pi_m(\unitn)$ ($-J\leq m\leq J$), along any direction $\unitn$ \cite{evrard_enhanced_2019}.

\begin{figure}
\includegraphics[
draft=false,scale=0.9,
trim={5mm 2mm 0 0.cm},
]{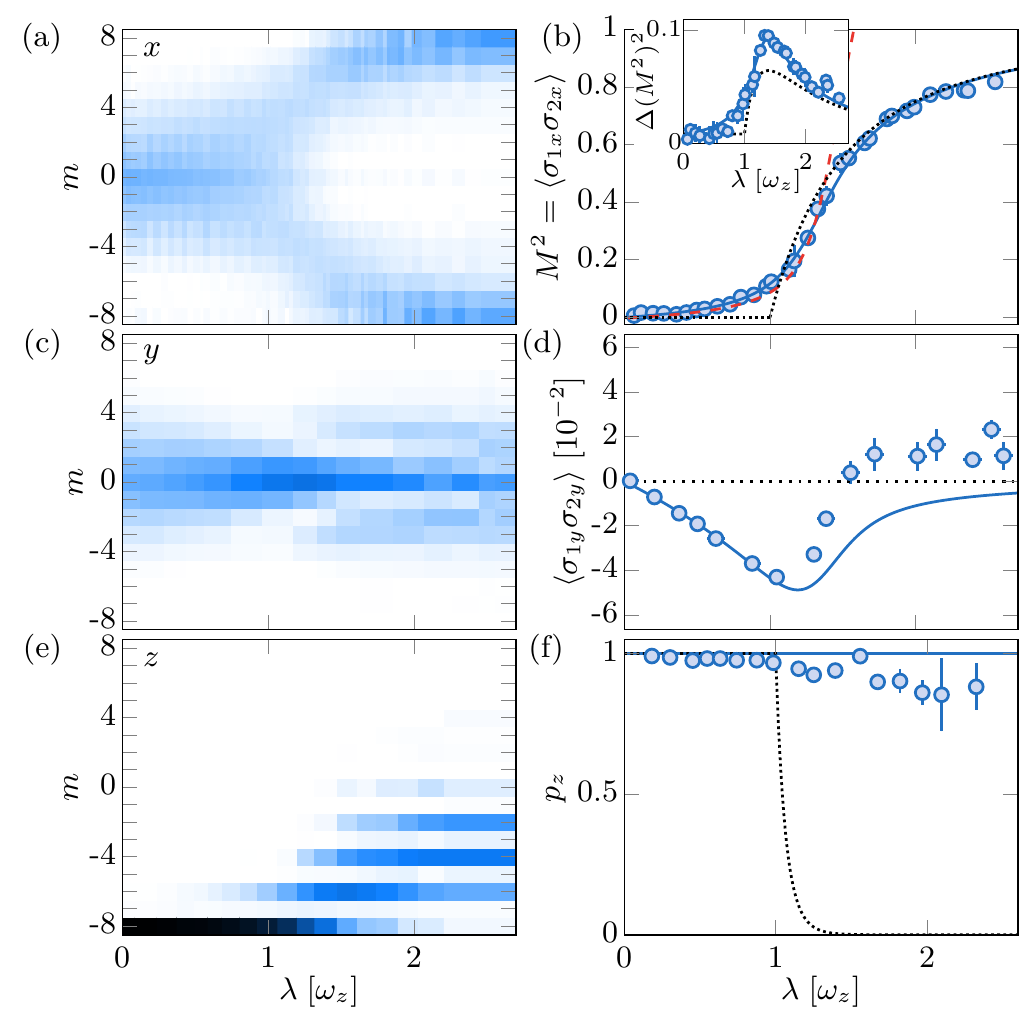}
\caption{
(a,c,e) Measured projection probabilities $\Pi_m(\unitn)$ for $\unitn=\unitx$, $\unity$ and $\unitz$ (a, c and e, respectively)  as a function of the interaction strength $\lambda$.
(b,d,f) Evolution of the spin pair correlator $\langle\sigma_{1x}\sigma_{2x}\rangle$ (b), its variance (inset of b), the correlator $\langle\sigma_{1y}\sigma_{2y}\rangle$ (d) and the mean parity $p_z$ (f). The solid blue (dotted black, dashed red) lines correspond to the  LMGm (classical mean-field model, critical Hamiltonian, respectively). No averaging in performed in (a). In other panels, all data are the average of about 5 independent measurements. In all figures error bars represent the 1-$\sigma$ statistical uncertainty.
\label{fig_ground_state}}
\end{figure}

We first investigate the properties of the ground state of the LMGm. We start with all atoms in the state $\ket{-J}_z$, which is the (paramagnetic) ground state for $\lambda=0$. We then slowly ramp the light coupling from zero up to a final value $\lambda$, using a linear ramp of speed $\dot\lambda\simeq0.015\,\omega_z^2$, for which we expect quasi-adiabatic evolution \cite{Note1}.  The measured spin projection probabilities $\Pi_m(\unitn)$ ($\unitn=\unitx,\unity,\unitz$) are shown in Fig.\,\ref{fig_ground_state}a,c,e. We first consider the occurrence of a ferromagnetic ground state, that would exhibit a non-zero order parameter $\langle\sigma_{1x}\rangle$. We show in Fig.\,\ref{fig_ground_state}a the single-shot projections $\Pi_m(\unitx)$ measured for various couplings $\lambda$. For small $\lambda$, we find a single-peak distribution centered on 0, consistent with the state $\ket{-J}_z$ projected along $\unitx$. For $\lambda\gtrsim\omega_z$, we observe a bifurcation towards a double-peak distribution, consistent with population of the two broken-symmetry states, each with an order parameter $\langle\sigma_{1x}\rangle\neq 0$. As the distributions remain globally symmetric,  the system does not seem to choose a single broken state.  Our measurement being averaged over many atoms, we cannot exclude a situation with almost half of the atoms in each broken state, e.g. organized in unresolved spin domains. This scenario is invalidated by a direct measurement of the mean parity $p_z\equiv\langle P_z\rangle$, that remains close to unity for all interaction strengths (see Fig.\,\ref{fig_ground_state}f).
Such an absence of spontaneous symmetry breaking is, in fact, expected for a finite-size system, whose ground state remains non-degenerate, as discussed later in this Letter. 

We now characterize the thermodynamic properties that are independent of  the symmetry breaking itself. We probe ferromagnetic spin correlations, i.e. the relative alignment between spins along $\unitx$ quantified by the correlator $M^2\equiv\langle\sigma_{1x}\sigma_{2x}\rangle$ \cite{botet_size_1982}.
We compute it from the second moment of the measured probabilities $\Pi_m(\unitx)$, using $N+N(N-1)\langle\sigma_{1\unitn}\sigma_{2\unitn}\rangle=4\langle J_{\unitn}^2\rangle$ \cite{ulam-orgikh_spin_2001}. As shown in Fig.\,\ref{fig_ground_state}b, the measurements agree well with the LMGm. The smooth increase of $M^2$ as a function of $\lambda$ is consistent with a crossover between para- and ferromagnetic behaviors. 

We also investigated signatures of the phase transition in our finite-size system.  First, we measure an increase of fluctuations of the ferromagnetic correlator $\Delta (M^2)\equiv \Delta(J_x^2)/[J(J-\tfrac{1}{2})]^2$ around the critical point $\lambda=\omega_z$ (see inset of Fig.\,\ref{fig_ground_state}b) -- a generic feature of continuous phase transitions \cite{LANDAU1980446}. More importantly, quantum phase transitions are also associated with the onset of entanglement in the critical region \cite{sachdev2011quantum}.  A priori, probing quantum entanglement  requires partitioning the electronic spin $J$, which is forbidden at low energy, but could, in principle, be achieved using coherent optical transitions $J\rightarrow J'$  \cite{guhne_entanglement_2009,killoran_extracting_2014-1}. Yet, we can indirectly probe entanglement in our system via spin projection correlations. Indeed, separable states which are symmetric upon exchange satisfy $\langle\sigma_{1\unitn}\sigma_{2\unitn}\rangle=\langle\sigma_{1\unitn}\rangle^2$ for all projection directions $\unitn$, and thus can only exhibit positive correlators 
 \cite{ulam-orgikh_spin_2001,vidal_concurrence_2006-1}. As shown in  Fig.\,\ref{fig_ground_state}c,d, we measure the correlator $\langle \sigma_{1y} \sigma_{2y}\rangle$  and show that it assumes negative values in a broad range of interaction strengths \cite{Note4}, consistent with quantum entanglement and suggesting that the phase transition is  driven by quantum (rather than thermal)  fluctuations \cite{osterloh_scaling_2002,luoDeterministicEntanglementGeneration2017}.

We now characterize more closely the region around the transition point, where the phase transition singularity is smoothened into a quantum critical behavior.  We focus on the variation of the ferromagnetic correlator $M^2$ with the coupling $\lambda$ (see Fig.\,\ref{fig_ground_state}b). Our measurements are close to  mean-field theory -- valid in the TL -- for most values of $\lambda$, except around  $\lambda=\omega_z$ \cite{landau_electrodynamics_1984,Note1}. In the critical regime, the leading finite-size correction can be simply formulated, as the quantum ground state remains close to the coherent state $\ket{-J}_z$, such that the operators $J_x$ and $J_y$ are almost canonically conjugated variables, with $[J_x,J_y]=\I J_z\simeq -\I J$ \cite{holsteinFieldDependenceIntrinsic1940}.
This approximation leads to a low-energy `critical' Hamiltonian \cite{ulyanov_new_1992,Note1}
\begin{equation}\label{eq_Hc}
\frac{H_c}{\hbar\omega_z}=-\left(J+\frac{1}{2}\right)+\frac{1}{J^{1/3}}\left(\frac{P^2}{2}-\frac{\epsilon X^2}{2}+\frac{X^4}{8}\right)
\end{equation}
describing the dynamics of a massive particle in a harmonic plus quartic potential, 
where $\epsilon=J^{2/3}(\lambda/\omega_z-1)$, $X=J^{-2/3}J_x$ and $P=-J^{-1/3}J_y$.  This description matches the textbook  Landau picture of a second-order phase transition, evolving from single- to double-well potentials  when crossing the critical point $\epsilon=0$ \cite{landau1937theory}. As plotted in Fig.\,\ref{fig_ground_state}b, the universal Hamiltonian (\ref{eq_Hc}) is sufficient to account well for the measured deviations to the TL around $\lambda=\omega_z$ \cite{Note6}.

\begin{figure}
\includegraphics[
draft=false,scale=0.9,
trim={3mm 6mm 0 0.cm},
]{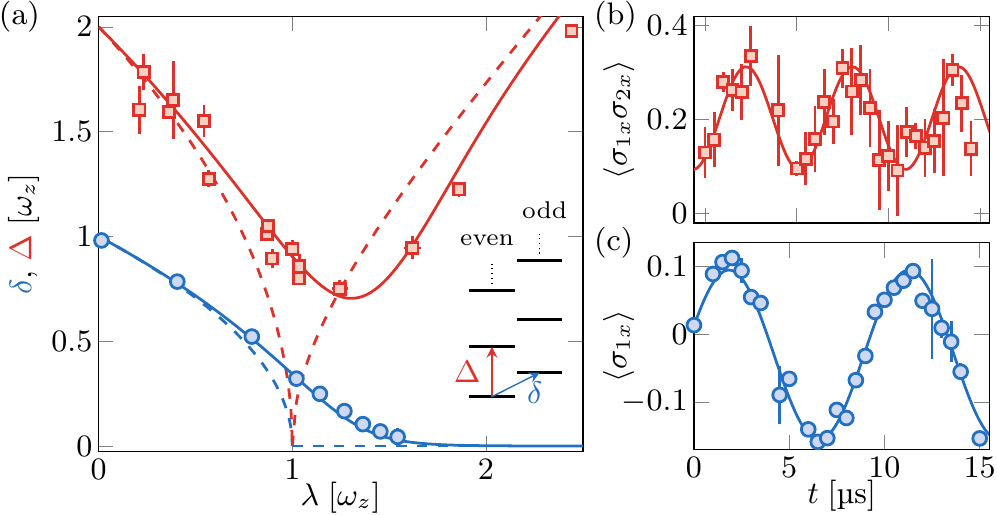}
\vspace{-0mm}
\caption{
(a) Parity gap $\delta$ between even- and odd-parity sectors and dynamical gap $\Delta$ between the ground and first even-parity states, as a function of the coupling $\lambda$. The solid (dashed) are the LMGm (mean-field) predictions. (a, inset) Energy level scheme of the 6 lowest eigenstates for $\lambda=0.5\,\omega_z$ (b) Breathing mode oscillation performed for $\lambda=1.04(2)\,\omega_z$. The solid line is a sine fit of frequency $\Delta$.  (c) Dipole mode oscillation performed for $\lambda=0.79(2)\,\omega_z$. The solid line is a sine fit of frequency $\delta$.
\label{fig_gap}}
\end{figure} 

We now extend our study to the dynamics of the system by measuring the energy gap of low-lying excitations. Due to the  $\mathbb{Z}_2$ parity symmetry of the LMGm, the eigenstates can be divided into two sectors of even and odd parity. The low-energy dynamics is then governed by two energy gaps, namely the `parity' gap $\hbar\delta$ between opposite-parity ground states and the `dynamical' gap $\hbar\Delta$ between the lowest two energy levels of even parity. In the effective potential picture, these gaps correspond to the oscillation frequencies of the dipole ($\delta$) and breathing ($\Delta$) modes. To excite the breathing mode for a given coupling $\lambda$, we simply increase the ramp speed $\dot\lambda$ used for the state preparation, leading to diabatic population of the first excited state of even parity, while keeping the higher states almost unpopulated. We then measure the time evolution of the second moment $\langle \sigma_{1x}\sigma_{2x}\rangle$, and extract its oscillation frequency $\Delta$ (see Fig.\;\ref{fig_gap}b). To excite the dipole mode, we first prepare the ground state for a given coupling $\lambda$, and apply a parity-breaking perturbation using a pulse of magnetic field along $x$ of duration $t\simeq\SI{3}{\micro s}$, coupling the ground state to the odd parity sector. The amplitude is chosen small enough to only populate the even- and odd-parity ground states, and the first moment $\langle \sigma_{1x}\rangle$ oscillation frequency, $\delta$, is extracted (see Fig.\;\ref{fig_gap}c). 

The measured parity and dynamical gaps, reported in Fig.\;\ref{fig_gap}a, agree well with the LMGm. The dynamical gap $\Delta$ exhibits a minimum around the critical point, reminiscent of the closing of the gap in the TL at the phase transition point. The parity gap $\delta$ decreases when increasing the coupling $\lambda$. In the paramagnetic phase $\lambda\lesssim0.5\,\omega_z$, the dynamical gap $\Delta$  remains about twice the parity gap $\delta$, consistent with a picture of non-interacting excitation quanta \cite{holsteinFieldDependenceIntrinsic1940,dusuel_finite-size_2004}. At the critical point, the measured dynamical gap $\Delta=0.91(5)\,\omega_z$ significantly exceeds twice the parity gap $\delta=0.33(1)\,\omega_z$, as expected from particle dynamics in a quartic potential (see the critical Hamiltonian (\ref{eq_Hc}) for $\epsilon=0$). This non-harmonic behavior illustrates the generic behavior of quantum critical systems, whose low-energy spectra cannot be simply reduced to non-interacting excitation quanta  \cite{sachdev2011quantum}. The gap value for $\lambda=\omega_z$ is also consistent with the leading finite-size correction to mean field $\Delta/\omega_z\simeq1.78/J^{1/3}=0.89$, valid for $J\gg1$ \cite{botet_size_1982,dusuel_continuous_2005-1,Note7}.

\begin{figure}
\includegraphics[
draft=false,scale=0.9,trim={3mm 3mm 0 0.cm},
]{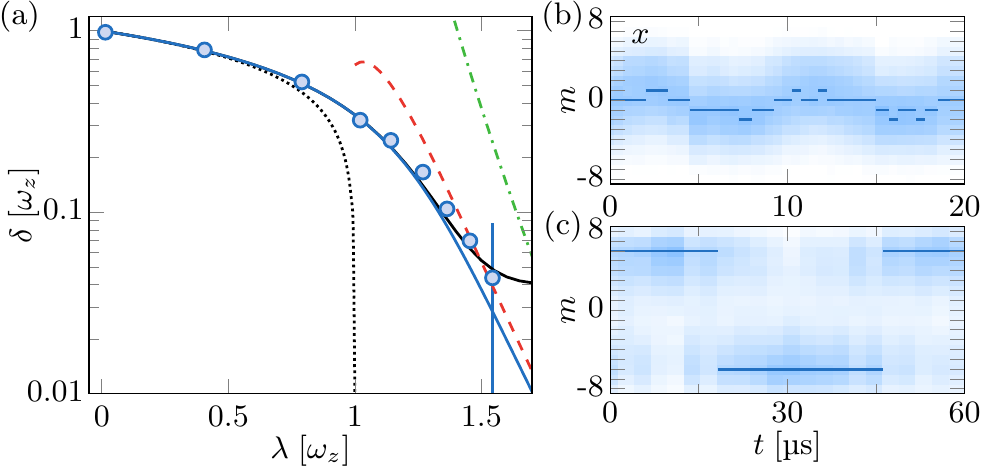}
\caption{
(a)  Parity gap $\delta$ as a function of  $\lambda$ (blue dots), compared with  the LMGm (blue line), mean-field theory (black dotted line), semi-classical tunneling (red dashed line) and perturbation theories (green dash-dotted line). The solid black line is the mean value of $\delta$ expected from the LMGm and averaged over  magnetic field fluctuations.
(b,c) Time evolution of the projection probabilities $\Pi_m(\unitx)$ during a dipole mode oscillation for $\lambda=0.79(2)\,\omega_z$ (b) and $\lambda=1.36(2)\,\omega_z$ (c). The most probable projection $m^*$ is plotted as a blue line.    
\label{fig_parity_gap}}
\end{figure}

We now focus on the dipole oscillation measurements in the ferromagnetic phase. In Fig.\;\ref{fig_parity_gap}a we plot the parity gap variation on a  logarithmic scale, showing a fast decrease  for $\lambda\gtrsim\omega_z$. The even- and odd-parity ground states thus become almost degenerate in the ferromagnetic phase, a behavior reminiscent of the exact double degeneracy expected in the TL for $\lambda>\omega_z$. We show in  Fig.\;\ref{fig_parity_gap}b,c the time evolution of the probability distributions $\Pi_m(\unitx)$ during the dipole oscillation, in the paramagnetic (b) and ferromagnetic (c) phases. In the paramagnetic phase the distributions always exhibit a single peak, whose center smoothly oscillates around zero. On the contrary, in the ferromagnetic phase  the distributions exhibit two peaks at positive/negative large-$|m|$ values, and the dynamics  consists in an oscillation between the peak weights, without significantly populating  small-$|m|$ states. This qualitatively different behavior is well illustrated by the evolution of the most probable projection $m^*$, which only takes two possible values $m^*=\pm6$ during the evolution shown in Fig.\;\ref{fig_parity_gap}c. These maximal projection values are close to the collective spin projections $\langle J_x\rangle=\pm5.4(5)$ of the two mean-field broken-symmetry states  for $\lambda=1.36(2)\,\omega_z$. Such a dynamics can be interpreted as a `macroscopic' quantum tunneling regime between broken states -- a phenomenon studied extensively in large-spin molecules \cite{gatteschi_quantum_2003,owerre_macroscopic_2015,friedman_macroscopic_1996-1,thomas_macroscopic_1996} and SQUID systems \cite{friedman_quantum_2000-1,wal_quantum_2000,makhlin_quantum-state_2001}. Deep in the ferromagnetic phase, the dipole frequencies are consistent with the semi-classical theory of quantum tunneling \cite{enz_spin_1986-1,scharf_tunnelling_1987,zaslavskii_spin_1990}. In the limit $\lambda\gg\omega_z$, perturbation theory provides a simple picture of this behavior: the two broken states $\ket{\pm J}_x$ being coupled by the $z$ field via a $2J$-order process lead to a high power-law scaling $\delta/\omega_z\propto(\omega_z/\lambda)^{2J-1}$. For values $\lambda\gtrsim1.5\,\omega_z$, the oscillation contrast decreases and the measured frequency deviates from theory, which we attribute to residual magnetic field fluctuations along $x$ (r.m.s. width $\sigma_B=\SI{0.4}{mG}$), inducing an offset between the two wells that exceeds the tunnel coupling.

\begin{figure}
\includegraphics[
draft=false,scale=0.9,trim={3mm 3mm 0 0.cm},
]{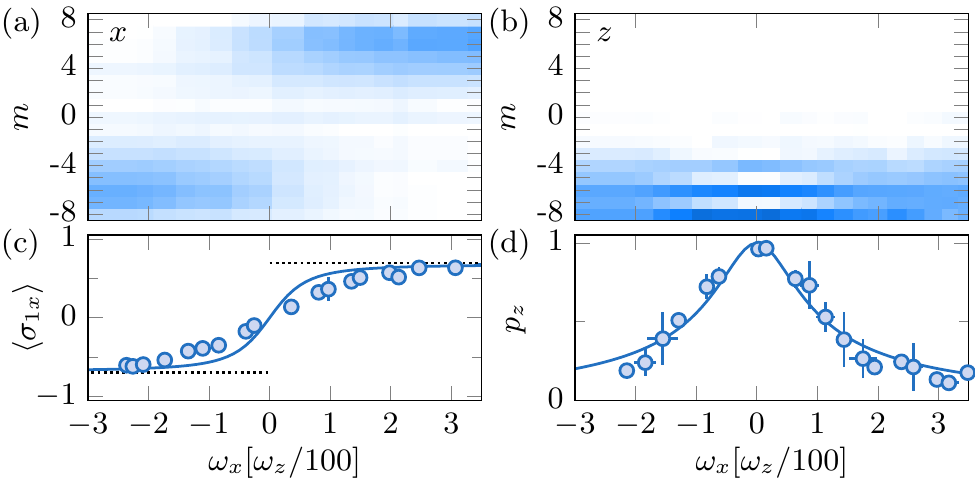}
\caption{
(a,b)  Projection probabilities $\Pi_m(\unitx)$ (a) and $\Pi_m(\unitz)$ (b) in the ground state as a function of  $\omega_x$, for   $\lambda=1.40(3)\,\omega_z$. (c,d) Order parameter $\langle\sigma_{1x}\rangle$ and mean parity $p_z$ computed from (a,b), and compared to the LMGm  (solid lines) and the  mean-field order parameter values (dotted lines).
\label{fig_parity_breaking}}
\end{figure} 

We finally investigate the controlled breaking of parity symmetry by a static magnetic field applied along $x$, which adds a Zeeman coupling $-\hbar\omega_x J_x$ mixing the two parity sectors. As shown in Fig.\,\ref{fig_parity_gap}, this field simultaneously induces  a  finite order parameter $\langle \sigma_{1x}\rangle$ and a reduction of the mean parity $p_z$. For large fields, the order parameter reaches a plateau consistent with the mean-field prediction $\langle \sigma_{1x}\rangle=\sgn(\omega_x)\sqrt{1-(\omega_z/\lambda)^2}$. This behavior coincides with a cancellation of  the mean parity $p_z$, illustrating the direct link between broken symmetry and non-zero order parameter \cite{Note1}.   Besides the controlled symmetry breaking discussed above, spontaneous symmetry breaking also occurs in our system  when preparing the ground state in the ferromagnetic phase, using very slow ramps of the light coupling $\dot\lambda\simeq10^{-3}\omega_z^2$. We find that the sign of the spontaneous order parameter $\langle \sigma_{1x}\rangle$ is directly related to the sign of the shot-to-shot magnetic field fluctuation, which is independently recorded. However, we found no signature of more complex symmetry-breaking mechanisms, e.g. induced by spin-dependent interactions between different atoms, as we did not observe a significant reduction of  parity when increasing the atomic density (up to $\sim10^{14}\,\si{cm^{-3}}$).

In conclusion, we studied the ground state and low-energy spectrum of the LMGm via the non-linear dynamics of the $J=8$ electronic spin of $^{162}$Dy atoms, observing a minimum of the energy gap around the transition point. A possible extension of this study would be the non-adiabatic crossing of the critical point, a problem related to quantum annealing \cite{bapst_quantum_2012} and Kibble-Zurek mechanism -- whose relevance for  infinitely coordinated systems is debated  \cite{caneva_adiabatic_2008-1,solinas_dynamical_2008-1,acevedo_new_2014,hwang_quantum_2015-2,defenu_dynamical_2018-2}. 
In the ferromagnetic phase we have demonstrated the production of coherent superposition of broken-symmetry states \cite{cirac_quantum_1998-1}, which could be used for quantum-enhanced metrology \cite{pezze_quantum_2018-1}.
Our system is also well suited to investigate various spontaneous  symmetry breaking mechanisms at the microscopic level and their connection to decoherence \cite{lucamariniTwoqubitEntanglementDynamics2004,vanwezelIntrinsicLimitQuantum2005}. 

\begin{acknowledgments}
We thank J. Dalibard and P. Ribeiro for stimulating discussions and J. Beugnon, J. Dalibard and F. Gerbier for a careful reading of the manuscript. This work is supported by PSL University (MAFAG project) and European Union (ERC
UQUAM and TOPODY). 
\end{acknowledgments}


\begin{thebibliography}{9}%
\makeatletter
\providecommand \@ifxundefined [1]{%
 \@ifx{#1\undefined}
}%
\providecommand \@ifnum [1]{%
 \ifnum #1\expandafter \@firstoftwo
 \else \expandafter \@secondoftwo
 \fi
}%
\providecommand \@ifx [1]{%
 \ifx #1\expandafter \@firstoftwo
 \else \expandafter \@secondoftwo
 \fi
}%
\providecommand \natexlab [1]{#1}%
\providecommand \enquote  [1]{``#1''}%
\providecommand \bibnamefont  [1]{#1}%
\providecommand \bibfnamefont [1]{#1}%
\providecommand \citenamefont [1]{#1}%
\providecommand \href@noop [0]{\@secondoftwo}%
\providecommand \href [0]{\begingroup \@sanitize@url \@href}%
\providecommand \@href[1]{\@@startlink{#1}\@@href}%
\providecommand \@@href[1]{\endgroup#1\@@endlink}%
\providecommand \@sanitize@url [0]{\catcode `\\12\catcode `\$12\catcode
  `\&12\catcode `\#12\catcode `\^12\catcode `\_12\catcode `\%12\relax}%
\providecommand \@@startlink[1]{}%
\providecommand \@@endlink[0]{}%
\providecommand \url  [0]{\begingroup\@sanitize@url \@url }%
\providecommand \@url [1]{\endgroup\@href {#1}{\urlprefix }}%
\providecommand \urlprefix  [0]{URL }%
\providecommand \Eprint [0]{\href }%
\providecommand \doibase [0]{http://dx.doi.org/}%
\providecommand \selectlanguage [0]{\@gobble}%
\providecommand \bibinfo  [0]{\@secondoftwo}%
\providecommand \bibfield  [0]{\@secondoftwo}%
\providecommand \translation [1]{[#1]}%
\providecommand \BibitemOpen [0]{}%
\providecommand \bibitemStop [0]{}%
\providecommand \bibitemNoStop [0]{.\EOS\space}%
\providecommand \EOS [0]{\spacefactor3000\relax}%
\providecommand \BibitemShut  [1]{\csname bibitem#1\endcsname}%
\let\auto@bib@innerbib\@empty
\bibitem [{\citenamefont {Ulyanov}\ and\ \citenamefont
  {Zaslavskii}(1992)}]{ulyanov_new_1992}%
  \BibitemOpen
  \bibfield  {author} {\bibinfo {author} {\bibfnamefont {V.~V.}\ \bibnamefont
  {Ulyanov}}\ and\ \bibinfo {author} {\bibfnamefont {O.~B.}\ \bibnamefont
  {Zaslavskii}},\ }\href {\doibase 10.1016/0370-1573(92)90158-V} {\bibfield
  {journal} {\bibinfo  {journal} {Physics Reports}\ }\textbf {\bibinfo {volume}
  {216}},\ \bibinfo {pages} {179} (\bibinfo {year} {1992})}\BibitemShut
  {NoStop}%
\bibitem [{\citenamefont {Holstein}\ and\ \citenamefont
  {Primakoff}(1940)}]{holsteinFieldDependenceIntrinsic1940}%
  \BibitemOpen
  \bibfield  {author} {\bibinfo {author} {\bibfnamefont {T.}~\bibnamefont
  {Holstein}}\ and\ \bibinfo {author} {\bibfnamefont {H.}~\bibnamefont
  {Primakoff}},\ }\href {\doibase 10.1103/PhysRev.58.1098} {\bibfield
  {journal} {\bibinfo  {journal} {Phys. Rev.}\ }\textbf {\bibinfo {volume}
  {58}},\ \bibinfo {pages} {1098} (\bibinfo {year} {1940})}\BibitemShut
  {NoStop}%
\bibitem [{\citenamefont {Dusuel}\ and\ \citenamefont
  {Vidal}(2004)}]{dusuel_finite-size_2004}%
  \BibitemOpen
  \bibfield  {author} {\bibinfo {author} {\bibfnamefont {S.}~\bibnamefont
  {Dusuel}}\ and\ \bibinfo {author} {\bibfnamefont {J.}~\bibnamefont {Vidal}},\
  }\href {\doibase 10.1103/PhysRevLett.93.237204} {\bibfield  {journal}
  {\bibinfo  {journal} {Phys. Rev. Lett.}\ }\textbf {\bibinfo {volume} {93}},\
  \bibinfo {pages} {237204} (\bibinfo {year} {2004})}\BibitemShut {NoStop}%
\bibitem [{\citenamefont {Hioe}\ \emph {et~al.}(1978)\citenamefont {Hioe},
  \citenamefont {MacMillen},\ and\ \citenamefont
  {Montroll}}]{hioeQuantumTheoryAnharmonic1978}%
  \BibitemOpen
  \bibfield  {author} {\bibinfo {author} {\bibfnamefont {F.~T.}\ \bibnamefont
  {Hioe}}, \bibinfo {author} {\bibfnamefont {D.}~\bibnamefont {MacMillen}}, \
  and\ \bibinfo {author} {\bibfnamefont {E.~W.}\ \bibnamefont {Montroll}},\
  }\href@noop {} {\bibfield  {journal} {\bibinfo  {journal} {Phys. Rep.}\
  }\textbf {\bibinfo {volume} {43}},\ \bibinfo {pages} {305} (\bibinfo {year}
  {1978})}\BibitemShut {NoStop}%
\bibitem [{\citenamefont {Leyvraz}\ and\ \citenamefont
  {Heiss}(2005)}]{leyvraz_large-$n$_2005}%
  \BibitemOpen
  \bibfield  {author} {\bibinfo {author} {\bibfnamefont {F.}~\bibnamefont
  {Leyvraz}}\ and\ \bibinfo {author} {\bibfnamefont {W.~D.}\ \bibnamefont
  {Heiss}},\ }\href {\doibase 10.1103/PhysRevLett.95.050402} {\bibfield
  {journal} {\bibinfo  {journal} {Phys. Rev. Lett.}\ }\textbf {\bibinfo
  {volume} {95}},\ \bibinfo {pages} {050402} (\bibinfo {year}
  {2005})}\BibitemShut {NoStop}%
\bibitem [{\citenamefont {Sachdev}(2011)}]{sachdev2011quantum}%
  \BibitemOpen
  \bibfield  {author} {\bibinfo {author} {\bibfnamefont {S.}~\bibnamefont
  {Sachdev}},\ }\href@noop {} {\emph {\bibinfo {title} {Quantum Phase
  Transitions}}}\ (\bibinfo  {publisher} {{Cambridge University Press}},\
  \bibinfo {year} {2011})\BibitemShut {NoStop}%
\bibitem [{\citenamefont {Heyl}(2018)}]{heylDynamicalQuantumPhase2018}%
  \BibitemOpen
  \bibfield  {author} {\bibinfo {author} {\bibfnamefont {M.}~\bibnamefont
  {Heyl}},\ }\href {\doibase 10.1088/1361-6633/aaaf9a} {\bibfield  {journal}
  {\bibinfo  {journal} {Rep. Prog. Phys.}\ }\textbf {\bibinfo {volume} {81}},\
  \bibinfo {pages} {054001} (\bibinfo {year} {2018})}\BibitemShut {NoStop}%
\bibitem [{\citenamefont {Botet}\ \emph {et~al.}(1982)\citenamefont {Botet},
  \citenamefont {Jullien},\ and\ \citenamefont {Pfeuty}}]{botet_size_1982}%
  \BibitemOpen
  \bibfield  {author} {\bibinfo {author} {\bibfnamefont {R.}~\bibnamefont
  {Botet}}, \bibinfo {author} {\bibfnamefont {R.}~\bibnamefont {Jullien}}, \
  and\ \bibinfo {author} {\bibfnamefont {P.}~\bibnamefont {Pfeuty}},\ }\href
  {\doibase 10.1103/PhysRevLett.49.478} {\bibfield  {journal} {\bibinfo
  {journal} {Phys. Rev. Lett.}\ }\textbf {\bibinfo {volume} {49}},\ \bibinfo
  {pages} {478} (\bibinfo {year} {1982})}\BibitemShut {NoStop}%
\bibitem [{\citenamefont {Dusuel}\ and\ \citenamefont
  {Vidal}(2005)}]{dusuel_continuous_2005-1}%
  \BibitemOpen
  \bibfield  {author} {\bibinfo {author} {\bibfnamefont {S.}~\bibnamefont
  {Dusuel}}\ and\ \bibinfo {author} {\bibfnamefont {J.}~\bibnamefont {Vidal}},\
  }\href {\doibase 10.1103/PhysRevB.71.224420} {\bibfield  {journal} {\bibinfo
  {journal} {Phys. Rev. B}\ }\textbf {\bibinfo {volume} {71}},\ \bibinfo
  {pages} {224420} (\bibinfo {year} {2005})}\BibitemShut {NoStop}%
\end{thebibliography}%


\begin{thebibliography}{60}%
\makeatletter
\providecommand \@ifxundefined [1]{%
 \@ifx{#1\undefined}
}%
\providecommand \@ifnum [1]{%
 \ifnum #1\expandafter \@firstoftwo
 \else \expandafter \@secondoftwo
 \fi
}%
\providecommand \@ifx [1]{%
 \ifx #1\expandafter \@firstoftwo
 \else \expandafter \@secondoftwo
 \fi
}%
\providecommand \natexlab [1]{#1}%
\providecommand \enquote  [1]{``#1''}%
\providecommand \bibnamefont  [1]{#1}%
\providecommand \bibfnamefont [1]{#1}%
\providecommand \citenamefont [1]{#1}%
\providecommand \href@noop [0]{\@secondoftwo}%
\providecommand \href [0]{\begingroup \@sanitize@url \@href}%
\providecommand \@href[1]{\@@startlink{#1}\@@href}%
\providecommand \@@href[1]{\endgroup#1\@@endlink}%
\providecommand \@sanitize@url [0]{\catcode `\\12\catcode `\$12\catcode
  `\&12\catcode `\#12\catcode `\^12\catcode `\_12\catcode `\%12\relax}%
\providecommand \@@startlink[1]{}%
\providecommand \@@endlink[0]{}%
\providecommand \url  [0]{\begingroup\@sanitize@url \@url }%
\providecommand \@url [1]{\endgroup\@href {#1}{\urlprefix }}%
\providecommand \urlprefix  [0]{URL }%
\providecommand \Eprint [0]{\href }%
\providecommand \doibase [0]{http://dx.doi.org/}%
\providecommand \selectlanguage [0]{\@gobble}%
\providecommand \bibinfo  [0]{\@secondoftwo}%
\providecommand \bibfield  [0]{\@secondoftwo}%
\providecommand \translation [1]{[#1]}%
\providecommand \BibitemOpen [0]{}%
\providecommand \bibitemStop [0]{}%
\providecommand \bibitemNoStop [0]{.\EOS\space}%
\providecommand \EOS [0]{\spacefactor3000\relax}%
\providecommand \BibitemShut  [1]{\csname bibitem#1\endcsname}%
\let\auto@bib@innerbib\@empty
\bibitem [{\citenamefont {Sachdev}(2011)}]{sachdev2011quantum}%
  \BibitemOpen
  \bibfield  {author} {\bibinfo {author} {\bibfnamefont {S.}~\bibnamefont
  {Sachdev}},\ }\href@noop {} {\emph {\bibinfo {title} {Quantum Phase
  Transitions}}}\ (\bibinfo  {publisher} {{Cambridge University Press}},\
  \bibinfo {year} {2011})\BibitemShut {NoStop}%
\bibitem [{\citenamefont {Osterloh}\ \emph {et~al.}(2002)\citenamefont
  {Osterloh}, \citenamefont {Amico}, \citenamefont {Falci},\ and\ \citenamefont
  {Fazio}}]{osterloh_scaling_2002}%
  \BibitemOpen
  \bibfield  {author} {\bibinfo {author} {\bibfnamefont {A.}~\bibnamefont
  {Osterloh}}, \bibinfo {author} {\bibfnamefont {L.}~\bibnamefont {Amico}},
  \bibinfo {author} {\bibfnamefont {G.}~\bibnamefont {Falci}}, \ and\ \bibinfo
  {author} {\bibfnamefont {R.}~\bibnamefont {Fazio}},\ }\href {\doibase
  10.1038/416608a} {\bibfield  {journal} {\bibinfo  {journal} {Nature}\
  }\textbf {\bibinfo {volume} {416}},\ \bibinfo {pages} {608} (\bibinfo {year}
  {2002})}\BibitemShut {NoStop}%
\bibitem [{\citenamefont {Blatt}\ and\ \citenamefont
  {Roos}(2012)}]{blatt_quantum_2012}%
  \BibitemOpen
  \bibfield  {author} {\bibinfo {author} {\bibfnamefont {R.}~\bibnamefont
  {Blatt}}\ and\ \bibinfo {author} {\bibfnamefont {C.~F.}\ \bibnamefont
  {Roos}},\ }\href {\doibase 10.1038/nphys2252} {\bibfield  {journal} {\bibinfo
   {journal} {Nat. Phys.}\ }\textbf {\bibinfo {volume} {8}},\ \bibinfo {pages}
  {277} (\bibinfo {year} {2012})}\BibitemShut {NoStop}%
\bibitem [{\citenamefont {Gross}\ and\ \citenamefont
  {Bloch}(2017)}]{gross_quantum_2017-1}%
  \BibitemOpen
  \bibfield  {author} {\bibinfo {author} {\bibfnamefont {C.}~\bibnamefont
  {Gross}}\ and\ \bibinfo {author} {\bibfnamefont {I.}~\bibnamefont {Bloch}},\
  }\href {\doibase 10.1126/science.aal3837} {\bibfield  {journal} {\bibinfo
  {journal} {Science}\ }\textbf {\bibinfo {volume} {357}},\ \bibinfo {pages}
  {995} (\bibinfo {year} {2017})}\BibitemShut {NoStop}%
\bibitem [{\citenamefont {Saffman}(2016)}]{saffman_quantum_2016}%
  \BibitemOpen
  \bibfield  {author} {\bibinfo {author} {\bibfnamefont {M.}~\bibnamefont
  {Saffman}},\ }\href {\doibase 10.1088/0953-4075/49/20/202001} {\bibfield
  {journal} {\bibinfo  {journal} {J. Phys. B: At. Mol. Opt. Phys.}\ }\textbf
  {\bibinfo {volume} {49}},\ \bibinfo {pages} {202001} (\bibinfo {year}
  {2016})}\BibitemShut {NoStop}%
\bibitem [{\citenamefont {Ma}\ \emph {et~al.}(2011)\citenamefont {Ma},
  \citenamefont {Dakic}, \citenamefont {Naylor}, \citenamefont {Zeilinger},\
  and\ \citenamefont {Walther}}]{ma_quantum_2011-1}%
  \BibitemOpen
  \bibfield  {author} {\bibinfo {author} {\bibfnamefont {X.-s.}\ \bibnamefont
  {Ma}}, \bibinfo {author} {\bibfnamefont {B.}~\bibnamefont {Dakic}}, \bibinfo
  {author} {\bibfnamefont {W.}~\bibnamefont {Naylor}}, \bibinfo {author}
  {\bibfnamefont {A.}~\bibnamefont {Zeilinger}}, \ and\ \bibinfo {author}
  {\bibfnamefont {P.}~\bibnamefont {Walther}},\ }\href {\doibase
  10.1038/nphys1919} {\bibfield  {journal} {\bibinfo  {journal} {Nat. Phys.}\
  }\textbf {\bibinfo {volume} {7}},\ \bibinfo {pages} {399} (\bibinfo {year}
  {2011})}\BibitemShut {NoStop}%
\bibitem [{\citenamefont {Georgescu}\ \emph {et~al.}(2014)\citenamefont
  {Georgescu}, \citenamefont {Ashhab},\ and\ \citenamefont
  {Nori}}]{georgescu_quantum_2014-1}%
  \BibitemOpen
  \bibfield  {author} {\bibinfo {author} {\bibfnamefont {I.~M.}\ \bibnamefont
  {Georgescu}}, \bibinfo {author} {\bibfnamefont {S.}~\bibnamefont {Ashhab}}, \
  and\ \bibinfo {author} {\bibfnamefont {F.}~\bibnamefont {Nori}},\ }\href
  {\doibase 10.1103/RevModPhys.86.153} {\bibfield  {journal} {\bibinfo
  {journal} {Rev. Mod. Phys.}\ }\textbf {\bibinfo {volume} {86}},\ \bibinfo
  {pages} {153} (\bibinfo {year} {2014})}\BibitemShut {NoStop}%
\bibitem [{\citenamefont {Islam}\ \emph {et~al.}(2015)\citenamefont {Islam},
  \citenamefont {Ma}, \citenamefont {Preiss}, \citenamefont {Eric~Tai},
  \citenamefont {Lukin}, \citenamefont {Rispoli},\ and\ \citenamefont
  {Greiner}}]{islamMeasuringEntanglementEntropy2015}%
  \BibitemOpen
  \bibfield  {author} {\bibinfo {author} {\bibfnamefont {R.}~\bibnamefont
  {Islam}}, \bibinfo {author} {\bibfnamefont {R.}~\bibnamefont {Ma}}, \bibinfo
  {author} {\bibfnamefont {P.~M.}\ \bibnamefont {Preiss}}, \bibinfo {author}
  {\bibfnamefont {M.}~\bibnamefont {Eric~Tai}}, \bibinfo {author}
  {\bibfnamefont {A.}~\bibnamefont {Lukin}}, \bibinfo {author} {\bibfnamefont
  {M.}~\bibnamefont {Rispoli}}, \ and\ \bibinfo {author} {\bibfnamefont
  {M.}~\bibnamefont {Greiner}},\ }\href {\doibase 10.1038/nature15750}
  {\bibfield  {journal} {\bibinfo  {journal} {Nature}\ }\textbf {\bibinfo
  {volume} {528}},\ \bibinfo {pages} {77} (\bibinfo {year} {2015})}\BibitemShut
  {NoStop}%
\bibitem [{\citenamefont {Hilker}\ \emph {et~al.}(2017)\citenamefont {Hilker},
  \citenamefont {Salomon}, \citenamefont {Grusdt}, \citenamefont {Omran},
  \citenamefont {Boll}, \citenamefont {Demler}, \citenamefont {Bloch},\ and\
  \citenamefont {Gross}}]{hilkerRevealingHiddenAntiferromagnetic2017}%
  \BibitemOpen
  \bibfield  {author} {\bibinfo {author} {\bibfnamefont {T.~A.}\ \bibnamefont
  {Hilker}}, \bibinfo {author} {\bibfnamefont {G.}~\bibnamefont {Salomon}},
  \bibinfo {author} {\bibfnamefont {F.}~\bibnamefont {Grusdt}}, \bibinfo
  {author} {\bibfnamefont {A.}~\bibnamefont {Omran}}, \bibinfo {author}
  {\bibfnamefont {M.}~\bibnamefont {Boll}}, \bibinfo {author} {\bibfnamefont
  {E.}~\bibnamefont {Demler}}, \bibinfo {author} {\bibfnamefont
  {I.}~\bibnamefont {Bloch}}, \ and\ \bibinfo {author} {\bibfnamefont
  {C.}~\bibnamefont {Gross}},\ }\href {\doibase 10.1126/science.aam8990}
  {\bibfield  {journal} {\bibinfo  {journal} {Science}\ }\textbf {\bibinfo
  {volume} {357}},\ \bibinfo {pages} {484} (\bibinfo {year}
  {2017})}\BibitemShut {NoStop}%
\bibitem [{\citenamefont {Landau}(1937)}]{landau1937theory}%
  \BibitemOpen
  \bibfield  {author} {\bibinfo {author} {\bibfnamefont {L.~D.}\ \bibnamefont
  {Landau}},\ }\href@noop {} {\bibfield  {journal} {\bibinfo  {journal} {Ukr J
  Phys}\ }\textbf {\bibinfo {volume} {11}},\ \bibinfo {pages} {19} (\bibinfo
  {year} {1937})}\BibitemShut {NoStop}%
\bibitem [{\citenamefont {Bartlett}\ \emph {et~al.}(2007)\citenamefont
  {Bartlett}, \citenamefont {Rudolph},\ and\ \citenamefont
  {Spekkens}}]{bartlett_reference_2007}%
  \BibitemOpen
  \bibfield  {author} {\bibinfo {author} {\bibfnamefont {S.~D.}\ \bibnamefont
  {Bartlett}}, \bibinfo {author} {\bibfnamefont {T.}~\bibnamefont {Rudolph}}, \
  and\ \bibinfo {author} {\bibfnamefont {R.~W.}\ \bibnamefont {Spekkens}},\
  }\href {\doibase 10.1103/RevModPhys.79.555} {\bibfield  {journal} {\bibinfo
  {journal} {Rev. Mod. Phys.}\ }\textbf {\bibinfo {volume} {79}},\ \bibinfo
  {pages} {555} (\bibinfo {year} {2007})}\BibitemShut {NoStop}%
\bibitem [{\citenamefont {Anderson}(1984)}]{anderson1984basic}%
  \BibitemOpen
  \bibfield  {author} {\bibinfo {author} {\bibfnamefont {P.}~\bibnamefont
  {Anderson}},\ }\href@noop {} {\emph {\bibinfo {title} {Basic {{Notions}} of
  {{Condensed Matter Physics}}}}}\ (\bibinfo  {publisher}
  {{Benjamin/Cummings}},\ \bibinfo {year} {1984})\BibitemShut {NoStop}%
\bibitem [{\citenamefont {Lipkin}\ \emph {et~al.}(1965)\citenamefont {Lipkin},
  \citenamefont {Meshkov},\ and\ \citenamefont {Glick}}]{lipkin_validity_1965}%
  \BibitemOpen
  \bibfield  {author} {\bibinfo {author} {\bibfnamefont {H.~J.}\ \bibnamefont
  {Lipkin}}, \bibinfo {author} {\bibfnamefont {N.}~\bibnamefont {Meshkov}}, \
  and\ \bibinfo {author} {\bibfnamefont {A.~J.}\ \bibnamefont {Glick}},\ }\href
  {\doibase 10.1016/0029-5582(65)90862-X} {\bibfield  {journal} {\bibinfo
  {journal} {Nucl. Phys.}\ }\textbf {\bibinfo {volume} {62}},\ \bibinfo {pages}
  {188} (\bibinfo {year} {1965})}\BibitemShut {NoStop}%
\bibitem [{\citenamefont {Cejnar}\ \emph {et~al.}(2010)\citenamefont {Cejnar},
  \citenamefont {Jolie},\ and\ \citenamefont {Casten}}]{cejnar_quantum_2010}%
  \BibitemOpen
  \bibfield  {author} {\bibinfo {author} {\bibfnamefont {P.}~\bibnamefont
  {Cejnar}}, \bibinfo {author} {\bibfnamefont {J.}~\bibnamefont {Jolie}}, \
  and\ \bibinfo {author} {\bibfnamefont {R.~F.}\ \bibnamefont {Casten}},\
  }\href {\doibase 10.1103/RevModPhys.82.2155} {\bibfield  {journal} {\bibinfo
  {journal} {Rev. Mod. Phys.}\ }\textbf {\bibinfo {volume} {82}},\ \bibinfo
  {pages} {2155} (\bibinfo {year} {2010})}\BibitemShut {NoStop}%
\bibitem [{\citenamefont {Gatteschi}\ and\ \citenamefont
  {Sessoli}(2003)}]{gatteschi_quantum_2003}%
  \BibitemOpen
  \bibfield  {author} {\bibinfo {author} {\bibfnamefont {D.}~\bibnamefont
  {Gatteschi}}\ and\ \bibinfo {author} {\bibfnamefont {R.}~\bibnamefont
  {Sessoli}},\ }\href {\doibase 10.1002/anie.200390099} {\bibfield  {journal}
  {\bibinfo  {journal} {Angew. Chem. Int. Ed.}\ }\textbf {\bibinfo {volume}
  {42}},\ \bibinfo {pages} {268} (\bibinfo {year} {2003})}\BibitemShut
  {NoStop}%
\bibitem [{\citenamefont {Lanyon}\ \emph {et~al.}(2011)\citenamefont {Lanyon},
  \citenamefont {Hempel}, \citenamefont {Nigg}, \citenamefont {M\"uller},
  \citenamefont {Gerritsma}, \citenamefont {Z\"ahringer}, \citenamefont
  {Schindler}, \citenamefont {Barreiro}, \citenamefont {Rambach}, \citenamefont
  {Kirchmair}, \citenamefont {Hennrich}, \citenamefont {Zoller}, \citenamefont
  {Blatt},\ and\ \citenamefont {Roos}}]{lanyon_universal_2011}%
  \BibitemOpen
  \bibfield  {author} {\bibinfo {author} {\bibfnamefont {B.~P.}\ \bibnamefont
  {Lanyon}}, \bibinfo {author} {\bibfnamefont {C.}~\bibnamefont {Hempel}},
  \bibinfo {author} {\bibfnamefont {D.}~\bibnamefont {Nigg}}, \bibinfo {author}
  {\bibfnamefont {M.}~\bibnamefont {M\"uller}}, \bibinfo {author}
  {\bibfnamefont {R.}~\bibnamefont {Gerritsma}}, \bibinfo {author}
  {\bibfnamefont {F.}~\bibnamefont {Z\"ahringer}}, \bibinfo {author}
  {\bibfnamefont {P.}~\bibnamefont {Schindler}}, \bibinfo {author}
  {\bibfnamefont {J.~T.}\ \bibnamefont {Barreiro}}, \bibinfo {author}
  {\bibfnamefont {M.}~\bibnamefont {Rambach}}, \bibinfo {author} {\bibfnamefont
  {G.}~\bibnamefont {Kirchmair}}, \bibinfo {author} {\bibfnamefont
  {M.}~\bibnamefont {Hennrich}}, \bibinfo {author} {\bibfnamefont
  {P.}~\bibnamefont {Zoller}}, \bibinfo {author} {\bibfnamefont
  {R.}~\bibnamefont {Blatt}}, \ and\ \bibinfo {author} {\bibfnamefont {C.~F.}\
  \bibnamefont {Roos}},\ }\href {\doibase 10.1126/science.1208001} {\bibfield
  {journal} {\bibinfo  {journal} {Science}\ }\textbf {\bibinfo {volume}
  {334}},\ \bibinfo {pages} {57} (\bibinfo {year} {2011})}\BibitemShut
  {NoStop}%
\bibitem [{\citenamefont {Islam}\ \emph {et~al.}(2011)\citenamefont {Islam},
  \citenamefont {Edwards}, \citenamefont {Kim}, \citenamefont {Korenblit},
  \citenamefont {Noh}, \citenamefont {Carmichael}, \citenamefont {Lin},
  \citenamefont {Duan}, \citenamefont {Joseph~Wang}, \citenamefont
  {Freericks},\ and\ \citenamefont {Monroe}}]{islam_onset_2011}%
  \BibitemOpen
  \bibfield  {author} {\bibinfo {author} {\bibfnamefont {R.}~\bibnamefont
  {Islam}}, \bibinfo {author} {\bibfnamefont {E.~E.}\ \bibnamefont {Edwards}},
  \bibinfo {author} {\bibfnamefont {K.}~\bibnamefont {Kim}}, \bibinfo {author}
  {\bibfnamefont {S.}~\bibnamefont {Korenblit}}, \bibinfo {author}
  {\bibfnamefont {C.}~\bibnamefont {Noh}}, \bibinfo {author} {\bibfnamefont
  {H.}~\bibnamefont {Carmichael}}, \bibinfo {author} {\bibfnamefont {G.-D.}\
  \bibnamefont {Lin}}, \bibinfo {author} {\bibfnamefont {L.-M.}\ \bibnamefont
  {Duan}}, \bibinfo {author} {\bibfnamefont {C.-C.}\ \bibnamefont
  {Joseph~Wang}}, \bibinfo {author} {\bibfnamefont {J.~K.}\ \bibnamefont
  {Freericks}}, \ and\ \bibinfo {author} {\bibfnamefont {C.}~\bibnamefont
  {Monroe}},\ }\href {\doibase 10.1038/ncomms1374} {\bibfield  {journal}
  {\bibinfo  {journal} {Nat. Commun.}\ }\textbf {\bibinfo {volume} {2}},\
  \bibinfo {pages} {377} (\bibinfo {year} {2011})}\BibitemShut {NoStop}%
\bibitem [{\citenamefont {Albiez}\ \emph {et~al.}(2005)\citenamefont {Albiez},
  \citenamefont {Gati}, \citenamefont {F\"olling}, \citenamefont {Hunsmann},
  \citenamefont {Cristiani},\ and\ \citenamefont
  {Oberthaler}}]{albiez_direct_2005}%
  \BibitemOpen
  \bibfield  {author} {\bibinfo {author} {\bibfnamefont {M.}~\bibnamefont
  {Albiez}}, \bibinfo {author} {\bibfnamefont {R.}~\bibnamefont {Gati}},
  \bibinfo {author} {\bibfnamefont {J.}~\bibnamefont {F\"olling}}, \bibinfo
  {author} {\bibfnamefont {S.}~\bibnamefont {Hunsmann}}, \bibinfo {author}
  {\bibfnamefont {M.}~\bibnamefont {Cristiani}}, \ and\ \bibinfo {author}
  {\bibfnamefont {M.~K.}\ \bibnamefont {Oberthaler}},\ }\href {\doibase
  10.1103/PhysRevLett.95.010402} {\bibfield  {journal} {\bibinfo  {journal}
  {Phys. Rev. Lett.}\ }\textbf {\bibinfo {volume} {95}},\ \bibinfo {pages}
  {010402} (\bibinfo {year} {2005})}\BibitemShut {NoStop}%
\bibitem [{\citenamefont {Levy}\ \emph {et~al.}(2007)\citenamefont {Levy},
  \citenamefont {Lahoud}, \citenamefont {Shomroni},\ and\ \citenamefont
  {Steinhauer}}]{levy_.c._2007}%
  \BibitemOpen
  \bibfield  {author} {\bibinfo {author} {\bibfnamefont {S.}~\bibnamefont
  {Levy}}, \bibinfo {author} {\bibfnamefont {E.}~\bibnamefont {Lahoud}},
  \bibinfo {author} {\bibfnamefont {I.}~\bibnamefont {Shomroni}}, \ and\
  \bibinfo {author} {\bibfnamefont {J.}~\bibnamefont {Steinhauer}},\ }\href
  {\doibase 10.1038/nature06186} {\bibfield  {journal} {\bibinfo  {journal}
  {Nature}\ }\textbf {\bibinfo {volume} {449}},\ \bibinfo {pages} {579}
  (\bibinfo {year} {2007})}\BibitemShut {NoStop}%
\bibitem [{\citenamefont {Trenkwalder}\ \emph {et~al.}(2016)\citenamefont
  {Trenkwalder}, \citenamefont {Spagnolli}, \citenamefont {Semeghini},
  \citenamefont {Coop}, \citenamefont {Landini}, \citenamefont {Castilho},
  \citenamefont {Pezz\`e}, \citenamefont {Modugno}, \citenamefont {Inguscio},
  \citenamefont {Smerzi},\ and\ \citenamefont
  {Fattori}}]{trenkwalder_quantum_2016}%
  \BibitemOpen
  \bibfield  {author} {\bibinfo {author} {\bibfnamefont {A.}~\bibnamefont
  {Trenkwalder}}, \bibinfo {author} {\bibfnamefont {G.}~\bibnamefont
  {Spagnolli}}, \bibinfo {author} {\bibfnamefont {G.}~\bibnamefont
  {Semeghini}}, \bibinfo {author} {\bibfnamefont {S.}~\bibnamefont {Coop}},
  \bibinfo {author} {\bibfnamefont {M.}~\bibnamefont {Landini}}, \bibinfo
  {author} {\bibfnamefont {P.}~\bibnamefont {Castilho}}, \bibinfo {author}
  {\bibfnamefont {L.}~\bibnamefont {Pezz\`e}}, \bibinfo {author} {\bibfnamefont
  {G.}~\bibnamefont {Modugno}}, \bibinfo {author} {\bibfnamefont
  {M.}~\bibnamefont {Inguscio}}, \bibinfo {author} {\bibfnamefont
  {A.}~\bibnamefont {Smerzi}}, \ and\ \bibinfo {author} {\bibfnamefont
  {M.}~\bibnamefont {Fattori}},\ }\href {\doibase 10.1038/nphys3743} {\bibfield
   {journal} {\bibinfo  {journal} {Nat. Phys.}\ }\textbf {\bibinfo {volume}
  {12}},\ \bibinfo {pages} {826} (\bibinfo {year} {2016})}\BibitemShut
  {NoStop}%
\bibitem [{\citenamefont {Landau}\ and\ \citenamefont
  {Lifshitz}(1958)}]{landau_quantum_1958}%
  \BibitemOpen
  \bibfield  {author} {\bibinfo {author} {\bibfnamefont {L.~D.}\ \bibnamefont
  {Landau}}\ and\ \bibinfo {author} {\bibfnamefont {E.}~\bibnamefont
  {Lifshitz}},\ }\href@noop {} {\emph {\bibinfo {title} {Quantum Mechanics,
  Non-Relativistic Theory}}}\ (\bibinfo  {publisher} {{London Pergamon
  Press}},\ \bibinfo {year} {1958})\BibitemShut {NoStop}%
\bibitem [{\citenamefont {Smith}\ \emph {et~al.}(2004)\citenamefont {Smith},
  \citenamefont {Chaudhury}, \citenamefont {Silberfarb}, \citenamefont
  {Deutsch},\ and\ \citenamefont {Jessen}}]{smith_continuous_2004-2}%
  \BibitemOpen
  \bibfield  {author} {\bibinfo {author} {\bibfnamefont {G.~A.}\ \bibnamefont
  {Smith}}, \bibinfo {author} {\bibfnamefont {S.}~\bibnamefont {Chaudhury}},
  \bibinfo {author} {\bibfnamefont {A.}~\bibnamefont {Silberfarb}}, \bibinfo
  {author} {\bibfnamefont {I.~H.}\ \bibnamefont {Deutsch}}, \ and\ \bibinfo
  {author} {\bibfnamefont {P.~S.}\ \bibnamefont {Jessen}},\ }\href {\doibase
  10.1103/PhysRevLett.93.163602} {\bibfield  {journal} {\bibinfo  {journal}
  {Phys. Rev. Lett.}\ }\textbf {\bibinfo {volume} {93}},\ \bibinfo {pages}
  {163602} (\bibinfo {year} {2004})}\BibitemShut {NoStop}%
\bibitem [{\citenamefont {Botet}\ \emph {et~al.}(1982)\citenamefont {Botet},
  \citenamefont {Jullien},\ and\ \citenamefont {Pfeuty}}]{botet_size_1982}%
  \BibitemOpen
  \bibfield  {author} {\bibinfo {author} {\bibfnamefont {R.}~\bibnamefont
  {Botet}}, \bibinfo {author} {\bibfnamefont {R.}~\bibnamefont {Jullien}}, \
  and\ \bibinfo {author} {\bibfnamefont {P.}~\bibnamefont {Pfeuty}},\ }\href
  {\doibase 10.1103/PhysRevLett.49.478} {\bibfield  {journal} {\bibinfo
  {journal} {Phys. Rev. Lett.}\ }\textbf {\bibinfo {volume} {49}},\ \bibinfo
  {pages} {478} (\bibinfo {year} {1982})}\BibitemShut {NoStop}%
\bibitem [{\citenamefont {Dusuel}\ and\ \citenamefont
  {Vidal}(2004)}]{dusuel_finite-size_2004}%
  \BibitemOpen
  \bibfield  {author} {\bibinfo {author} {\bibfnamefont {S.}~\bibnamefont
  {Dusuel}}\ and\ \bibinfo {author} {\bibfnamefont {J.}~\bibnamefont {Vidal}},\
  }\href {\doibase 10.1103/PhysRevLett.93.237204} {\bibfield  {journal}
  {\bibinfo  {journal} {Phys. Rev. Lett.}\ }\textbf {\bibinfo {volume} {93}},\
  \bibinfo {pages} {237204} (\bibinfo {year} {2004})}\BibitemShut {NoStop}%
\bibitem [{\citenamefont {Campa}\ \emph {et~al.}(2014)\citenamefont {Campa},
  \citenamefont {Dauxois}, \citenamefont {Fanelli},\ and\ \citenamefont
  {Ruffo}}]{campa_physics_2014}%
  \BibitemOpen
  \bibfield  {author} {\bibinfo {author} {\bibfnamefont {A.}~\bibnamefont
  {Campa}}, \bibinfo {author} {\bibfnamefont {T.}~\bibnamefont {Dauxois}},
  \bibinfo {author} {\bibfnamefont {D.}~\bibnamefont {Fanelli}}, \ and\
  \bibinfo {author} {\bibfnamefont {S.}~\bibnamefont {Ruffo}},\ }\href@noop {}
  {\emph {\bibinfo {title} {Physics of {{Long}}-{{Range Interacting
  Systems}}}}}\ (\bibinfo  {publisher} {{Oxford University Press}},\ \bibinfo
  {year} {2014})\BibitemShut {NoStop}%
\bibitem [{\citenamefont {Evrard}\ \emph {et~al.}(2019)\citenamefont {Evrard},
  \citenamefont {Makhalov}, \citenamefont {Chalopin}, \citenamefont
  {Sidorenkov}, \citenamefont {Dalibard}, \citenamefont {Lopes},\ and\
  \citenamefont {Nascimbene}}]{evrard_enhanced_2019}%
  \BibitemOpen
  \bibfield  {author} {\bibinfo {author} {\bibfnamefont {A.}~\bibnamefont
  {Evrard}}, \bibinfo {author} {\bibfnamefont {V.}~\bibnamefont {Makhalov}},
  \bibinfo {author} {\bibfnamefont {T.}~\bibnamefont {Chalopin}}, \bibinfo
  {author} {\bibfnamefont {L.~A.}\ \bibnamefont {Sidorenkov}}, \bibinfo
  {author} {\bibfnamefont {J.}~\bibnamefont {Dalibard}}, \bibinfo {author}
  {\bibfnamefont {R.}~\bibnamefont {Lopes}}, \ and\ \bibinfo {author}
  {\bibfnamefont {S.}~\bibnamefont {Nascimbene}},\ }\href {\doibase
  10.1103/PhysRevLett.122.173601} {\bibfield  {journal} {\bibinfo  {journal}
  {Phys. Rev. Lett.}\ }\textbf {\bibinfo {volume} {122}},\ \bibinfo {pages}
  {173601} (\bibinfo {year} {2019})}\BibitemShut {NoStop}%
\bibitem [{Note1()}]{Note1}%
  \BibitemOpen
  \bibinfo {note} {See Supplemental Material for a study of adiabaticity
  requirements for the ground state preparation, eigenstate spectroscopy at
  higher energy, a description of the classical mean-field treatment, a
  derivation of the critical Hamiltonian, and an extensive discussion of parity
  symmetry breaking by an external magnetic field.}\BibitemShut {Stop}%
\bibitem [{\citenamefont {{Ulam-Orgikh}}\ and\ \citenamefont
  {Kitagawa}(2001)}]{ulam-orgikh_spin_2001}%
  \BibitemOpen
  \bibfield  {author} {\bibinfo {author} {\bibfnamefont {D.}~\bibnamefont
  {{Ulam-Orgikh}}}\ and\ \bibinfo {author} {\bibfnamefont {M.}~\bibnamefont
  {Kitagawa}},\ }\href {\doibase 10.1103/PhysRevA.64.052106} {\bibfield
  {journal} {\bibinfo  {journal} {Phys. Rev. A}\ }\textbf {\bibinfo {volume}
  {64}},\ \bibinfo {pages} {052106} (\bibinfo {year} {2001})}\BibitemShut
  {NoStop}%
\bibitem [{\citenamefont {Landau}\ and\ \citenamefont
  {Lifshitz}(1980)}]{LANDAU1980446}%
  \BibitemOpen
  \bibfield  {author} {\bibinfo {author} {\bibfnamefont {L.}~\bibnamefont
  {Landau}}\ and\ \bibinfo {author} {\bibfnamefont {E.}~\bibnamefont
  {Lifshitz}},\ }in\ \href {\doibase
  https://doi.org/10.1016/B978-0-08-057046-4.50021-X} {\emph {\bibinfo
  {booktitle} {Statistical {{Physics}}}}}\ (\bibinfo  {publisher}
  {{Butterworth-Heinemann}},\ \bibinfo {address} {Oxford},\ \bibinfo {year}
  {1980})\ \bibinfo {edition} {third edition}\ ed.,\ pp.\ \bibinfo {pages}
  {446--516}\BibitemShut {NoStop}%
\bibitem [{\citenamefont {G\"uhne}\ and\ \citenamefont
  {T\'oth}(2009)}]{guhne_entanglement_2009}%
  \BibitemOpen
  \bibfield  {author} {\bibinfo {author} {\bibfnamefont {O.}~\bibnamefont
  {G\"uhne}}\ and\ \bibinfo {author} {\bibfnamefont {G.}~\bibnamefont
  {T\'oth}},\ }\href {\doibase 10.1016/j.physrep.2009.02.004} {\bibfield
  {journal} {\bibinfo  {journal} {Physics Reports}\ }\textbf {\bibinfo {volume}
  {474}},\ \bibinfo {pages} {1} (\bibinfo {year} {2009})}\BibitemShut {NoStop}%
\bibitem [{\citenamefont {Killoran}\ \emph {et~al.}(2014)\citenamefont
  {Killoran}, \citenamefont {Cramer},\ and\ \citenamefont
  {Plenio}}]{killoran_extracting_2014-1}%
  \BibitemOpen
  \bibfield  {author} {\bibinfo {author} {\bibfnamefont {N.}~\bibnamefont
  {Killoran}}, \bibinfo {author} {\bibfnamefont {M.}~\bibnamefont {Cramer}}, \
  and\ \bibinfo {author} {\bibfnamefont {M.~B.}\ \bibnamefont {Plenio}},\
  }\href {\doibase 10.1103/PhysRevLett.112.150501} {\bibfield  {journal}
  {\bibinfo  {journal} {Phys. Rev. Lett.}\ }\textbf {\bibinfo {volume} {112}},\
  \bibinfo {pages} {150501} (\bibinfo {year} {2014})}\BibitemShut {NoStop}%
\bibitem [{\citenamefont {Vidal}(2006)}]{vidal_concurrence_2006-1}%
  \BibitemOpen
  \bibfield  {author} {\bibinfo {author} {\bibfnamefont {J.}~\bibnamefont
  {Vidal}},\ }\href {\doibase 10.1103/PhysRevA.73.062318} {\bibfield  {journal}
  {\bibinfo  {journal} {Phys. Rev. A}\ }\textbf {\bibinfo {volume} {73}},\
  \bibinfo {pages} {062318} (\bibinfo {year} {2006})}\BibitemShut {NoStop}%
\bibitem [{Note4()}]{Note4}%
  \BibitemOpen
  \bibinfo {note} {The minimum value $\left < \sigma _{1y} \sigma _{2y}\right
  >=-0.040(5)$ for $\lambda \simeq \omega _z$ is about 60\% of the minimum
  correlation $\left <\sigma _{1\protect \mathaccentV {hat}05E{\protect \mathbf
  {n}}} \sigma _{2\protect \mathaccentV {hat}05E{\protect \mathbf {n}}}\right
  >=-1/(N-1)$ allowed for a set of $N$ spins symmetric upon exchange \cite
  {koashi_entangled_2000}.}\BibitemShut {Stop}%
\bibitem [{\citenamefont {Luo}\ \emph {et~al.}(2017)\citenamefont {Luo},
  \citenamefont {Zou}, \citenamefont {Wu}, \citenamefont {Liu}, \citenamefont
  {Han}, \citenamefont {Tey},\ and\ \citenamefont
  {You}}]{luoDeterministicEntanglementGeneration2017}%
  \BibitemOpen
  \bibfield  {author} {\bibinfo {author} {\bibfnamefont {X.-Y.}\ \bibnamefont
  {Luo}}, \bibinfo {author} {\bibfnamefont {Y.-Q.}\ \bibnamefont {Zou}},
  \bibinfo {author} {\bibfnamefont {L.-N.}\ \bibnamefont {Wu}}, \bibinfo
  {author} {\bibfnamefont {Q.}~\bibnamefont {Liu}}, \bibinfo {author}
  {\bibfnamefont {M.-F.}\ \bibnamefont {Han}}, \bibinfo {author} {\bibfnamefont
  {M.~K.}\ \bibnamefont {Tey}}, \ and\ \bibinfo {author} {\bibfnamefont
  {L.}~\bibnamefont {You}},\ }\href {\doibase 10.1126/science.aag1106}
  {\bibfield  {journal} {\bibinfo  {journal} {Science}\ }\textbf {\bibinfo
  {volume} {355}},\ \bibinfo {pages} {620} (\bibinfo {year}
  {2017})}\BibitemShut {NoStop}%
\bibitem [{\citenamefont {Landau}\ \emph {et~al.}(1984)\citenamefont {Landau},
  \citenamefont {Pitaevskii},\ and\ \citenamefont
  {Lifshitz}}]{landau_electrodynamics_1984}%
  \BibitemOpen
  \bibfield  {author} {\bibinfo {author} {\bibfnamefont {L.}~\bibnamefont
  {Landau}}, \bibinfo {author} {\bibfnamefont {L.}~\bibnamefont {Pitaevskii}},
  \ and\ \bibinfo {author} {\bibfnamefont {E.}~\bibnamefont {Lifshitz}},\
  }\href@noop {} {\emph {\bibinfo {title} {Electrodynamics of {{Continuous
  Media}}}}}\ (\bibinfo  {publisher} {{Butterworth-Heinemann}},\ \bibinfo
  {year} {1984})\BibitemShut {NoStop}%
\bibitem [{\citenamefont {Holstein}\ and\ \citenamefont
  {Primakoff}(1940)}]{holsteinFieldDependenceIntrinsic1940}%
  \BibitemOpen
  \bibfield  {author} {\bibinfo {author} {\bibfnamefont {T.}~\bibnamefont
  {Holstein}}\ and\ \bibinfo {author} {\bibfnamefont {H.}~\bibnamefont
  {Primakoff}},\ }\href {\doibase 10.1103/PhysRev.58.1098} {\bibfield
  {journal} {\bibinfo  {journal} {Phys. Rev.}\ }\textbf {\bibinfo {volume}
  {58}},\ \bibinfo {pages} {1098} (\bibinfo {year} {1940})}\BibitemShut
  {NoStop}%
\bibitem [{\citenamefont {Ulyanov}\ and\ \citenamefont
  {Zaslavskii}(1992)}]{ulyanov_new_1992}%
  \BibitemOpen
  \bibfield  {author} {\bibinfo {author} {\bibfnamefont {V.~V.}\ \bibnamefont
  {Ulyanov}}\ and\ \bibinfo {author} {\bibfnamefont {O.~B.}\ \bibnamefont
  {Zaslavskii}},\ }\href {\doibase 10.1016/0370-1573(92)90158-V} {\bibfield
  {journal} {\bibinfo  {journal} {Physics Reports}\ }\textbf {\bibinfo {volume}
  {216}},\ \bibinfo {pages} {179} (\bibinfo {year} {1992})}\BibitemShut
  {NoStop}%
\bibitem [{Note6()}]{Note6}%
  \BibitemOpen
  \bibinfo {note} {It becomes inaccurate in the deep ferromagnetic phase (for
  $\lambda \gtrsim 1.5\protect \tmspace +\thinmuskip {.1667em}\omega _z$, see
  \cite {Note1}), where the precise shape of the double-well potential is
  required to describe quantitatively quantum tunneling effects \cite
  {ulyanov_new_1992}.}\BibitemShut {Stop}%
\bibitem [{\citenamefont {Dusuel}\ and\ \citenamefont
  {Vidal}(2005)}]{dusuel_continuous_2005-1}%
  \BibitemOpen
  \bibfield  {author} {\bibinfo {author} {\bibfnamefont {S.}~\bibnamefont
  {Dusuel}}\ and\ \bibinfo {author} {\bibfnamefont {J.}~\bibnamefont {Vidal}},\
  }\href {\doibase 10.1103/PhysRevB.71.224420} {\bibfield  {journal} {\bibinfo
  {journal} {Phys. Rev. B}\ }\textbf {\bibinfo {volume} {71}},\ \bibinfo
  {pages} {224420} (\bibinfo {year} {2005})}\BibitemShut {NoStop}%
\bibitem [{Note7()}]{Note7}%
  \BibitemOpen
  \bibinfo {note} {We present in the Supplementary Material measurements of
  higher energy levels, that confirm this picture on a larger energy
  range.}\BibitemShut {Stop}%
\bibitem [{\citenamefont {Owerre}\ and\ \citenamefont
  {Paranjape}(2015)}]{owerre_macroscopic_2015}%
  \BibitemOpen
  \bibfield  {author} {\bibinfo {author} {\bibfnamefont {S.~A.}\ \bibnamefont
  {Owerre}}\ and\ \bibinfo {author} {\bibfnamefont {M.~B.}\ \bibnamefont
  {Paranjape}},\ }\href {\doibase 10.1016/j.physrep.2014.09.001} {\bibfield
  {journal} {\bibinfo  {journal} {Phys. Rep.}\ }\textbf {\bibinfo {volume}
  {546}},\ \bibinfo {pages} {1} (\bibinfo {year} {2015})}\BibitemShut {NoStop}%
\bibitem [{\citenamefont {Friedman}\ \emph {et~al.}(1996)\citenamefont
  {Friedman}, \citenamefont {Sarachik}, \citenamefont {Tejada},\ and\
  \citenamefont {Ziolo}}]{friedman_macroscopic_1996-1}%
  \BibitemOpen
  \bibfield  {author} {\bibinfo {author} {\bibfnamefont {J.~R.}\ \bibnamefont
  {Friedman}}, \bibinfo {author} {\bibfnamefont {M.~P.}\ \bibnamefont
  {Sarachik}}, \bibinfo {author} {\bibfnamefont {J.}~\bibnamefont {Tejada}}, \
  and\ \bibinfo {author} {\bibfnamefont {R.}~\bibnamefont {Ziolo}},\ }\href
  {\doibase 10.1103/PhysRevLett.76.3830} {\bibfield  {journal} {\bibinfo
  {journal} {Phys. Rev. Lett.}\ }\textbf {\bibinfo {volume} {76}},\ \bibinfo
  {pages} {3830} (\bibinfo {year} {1996})}\BibitemShut {NoStop}%
\bibitem [{\citenamefont {Thomas}\ \emph {et~al.}(1996)\citenamefont {Thomas},
  \citenamefont {Lionti}, \citenamefont {Ballou}, \citenamefont {Gatteschi},
  \citenamefont {Sessoli},\ and\ \citenamefont
  {Barbara}}]{thomas_macroscopic_1996}%
  \BibitemOpen
  \bibfield  {author} {\bibinfo {author} {\bibfnamefont {L.}~\bibnamefont
  {Thomas}}, \bibinfo {author} {\bibfnamefont {F.}~\bibnamefont {Lionti}},
  \bibinfo {author} {\bibfnamefont {R.}~\bibnamefont {Ballou}}, \bibinfo
  {author} {\bibfnamefont {D.}~\bibnamefont {Gatteschi}}, \bibinfo {author}
  {\bibfnamefont {R.}~\bibnamefont {Sessoli}}, \ and\ \bibinfo {author}
  {\bibfnamefont {B.}~\bibnamefont {Barbara}},\ }\href {\doibase
  10.1038/383145a0} {\bibfield  {journal} {\bibinfo  {journal} {Nature}\
  }\textbf {\bibinfo {volume} {383}},\ \bibinfo {pages} {145} (\bibinfo {year}
  {1996})}\BibitemShut {NoStop}%
\bibitem [{\citenamefont {Friedman}\ \emph {et~al.}(2000)\citenamefont
  {Friedman}, \citenamefont {Patel}, \citenamefont {Chen}, \citenamefont
  {Tolpygo},\ and\ \citenamefont {Lukens}}]{friedman_quantum_2000-1}%
  \BibitemOpen
  \bibfield  {author} {\bibinfo {author} {\bibfnamefont {J.~R.}\ \bibnamefont
  {Friedman}}, \bibinfo {author} {\bibfnamefont {V.}~\bibnamefont {Patel}},
  \bibinfo {author} {\bibfnamefont {W.}~\bibnamefont {Chen}}, \bibinfo {author}
  {\bibfnamefont {S.~K.}\ \bibnamefont {Tolpygo}}, \ and\ \bibinfo {author}
  {\bibfnamefont {J.~E.}\ \bibnamefont {Lukens}},\ }\href {\doibase
  10.1038/35017505} {\bibfield  {journal} {\bibinfo  {journal} {Nature}\
  }\textbf {\bibinfo {volume} {406}},\ \bibinfo {pages} {43} (\bibinfo {year}
  {2000})}\BibitemShut {NoStop}%
\bibitem [{\citenamefont {van~der Wal}\ \emph {et~al.}(2000)\citenamefont
  {van~der Wal}, \citenamefont {ter Haar}, \citenamefont {Wilhelm},
  \citenamefont {Schouten}, \citenamefont {Harmans}, \citenamefont {Orlando},
  \citenamefont {Lloyd},\ and\ \citenamefont {Mooij}}]{wal_quantum_2000}%
  \BibitemOpen
  \bibfield  {author} {\bibinfo {author} {\bibfnamefont {C.~H.}\ \bibnamefont
  {van~der Wal}}, \bibinfo {author} {\bibfnamefont {A.~C.~J.}\ \bibnamefont
  {ter Haar}}, \bibinfo {author} {\bibfnamefont {F.~K.}\ \bibnamefont
  {Wilhelm}}, \bibinfo {author} {\bibfnamefont {R.~N.}\ \bibnamefont
  {Schouten}}, \bibinfo {author} {\bibfnamefont {C.~J. P.~M.}\ \bibnamefont
  {Harmans}}, \bibinfo {author} {\bibfnamefont {T.~P.}\ \bibnamefont
  {Orlando}}, \bibinfo {author} {\bibfnamefont {S.}~\bibnamefont {Lloyd}}, \
  and\ \bibinfo {author} {\bibfnamefont {J.~E.}\ \bibnamefont {Mooij}},\ }\href
  {\doibase 10.1126/science.290.5492.773} {\bibfield  {journal} {\bibinfo
  {journal} {Science}\ }\textbf {\bibinfo {volume} {290}},\ \bibinfo {pages}
  {773} (\bibinfo {year} {2000})}\BibitemShut {NoStop}%
\bibitem [{\citenamefont {Makhlin}\ \emph {et~al.}(2001)\citenamefont
  {Makhlin}, \citenamefont {Sch\"on},\ and\ \citenamefont
  {Shnirman}}]{makhlin_quantum-state_2001}%
  \BibitemOpen
  \bibfield  {author} {\bibinfo {author} {\bibfnamefont {Y.}~\bibnamefont
  {Makhlin}}, \bibinfo {author} {\bibfnamefont {G.}~\bibnamefont {Sch\"on}}, \
  and\ \bibinfo {author} {\bibfnamefont {A.}~\bibnamefont {Shnirman}},\ }\href
  {\doibase 10.1103/RevModPhys.73.357} {\bibfield  {journal} {\bibinfo
  {journal} {Rev. Mod. Phys.}\ }\textbf {\bibinfo {volume} {73}},\ \bibinfo
  {pages} {357} (\bibinfo {year} {2001})}\BibitemShut {NoStop}%
\bibitem [{\citenamefont {Enz}\ and\ \citenamefont
  {Schilling}(1986)}]{enz_spin_1986-1}%
  \BibitemOpen
  \bibfield  {author} {\bibinfo {author} {\bibfnamefont {M.}~\bibnamefont
  {Enz}}\ and\ \bibinfo {author} {\bibfnamefont {R.}~\bibnamefont
  {Schilling}},\ }\href {\doibase 10.1088/0022-3719/19/11/014} {\bibfield
  {journal} {\bibinfo  {journal} {J. Phys. C: Solid State Phys.}\ }\textbf
  {\bibinfo {volume} {19}},\ \bibinfo {pages} {1765} (\bibinfo {year}
  {1986})}\BibitemShut {NoStop}%
\bibitem [{\citenamefont {Scharf}\ \emph {et~al.}(1987)\citenamefont {Scharf},
  \citenamefont {Wreszinski}, \citenamefont {{van}},\ and\ \citenamefont
  {Hemmen}}]{scharf_tunnelling_1987}%
  \BibitemOpen
  \bibfield  {author} {\bibinfo {author} {\bibfnamefont {G.}~\bibnamefont
  {Scharf}}, \bibinfo {author} {\bibfnamefont {W.~F.}\ \bibnamefont
  {Wreszinski}}, \bibinfo {author} {\bibnamefont {{van}}}, \ and\ \bibinfo
  {author} {\bibfnamefont {J.~L.}\ \bibnamefont {Hemmen}},\ }\href {\doibase
  10.1088/0305-4470/20/13/032} {\bibfield  {journal} {\bibinfo  {journal} {J.
  Phys. A: Math. Gen.}\ }\textbf {\bibinfo {volume} {20}},\ \bibinfo {pages}
  {4309} (\bibinfo {year} {1987})}\BibitemShut {NoStop}%
\bibitem [{\citenamefont {Zaslavskii}(1990)}]{zaslavskii_spin_1990}%
  \BibitemOpen
  \bibfield  {author} {\bibinfo {author} {\bibfnamefont {O.~B.}\ \bibnamefont
  {Zaslavskii}},\ }\href {\doibase 10.1016/0375-9601(90)90317-H} {\bibfield
  {journal} {\bibinfo  {journal} {Phys. Lett. A}\ }\textbf {\bibinfo {volume}
  {145}},\ \bibinfo {pages} {471} (\bibinfo {year} {1990})}\BibitemShut
  {NoStop}%
\bibitem [{\citenamefont {Bapst}\ and\ \citenamefont
  {Semerjian}(2012)}]{bapst_quantum_2012}%
  \BibitemOpen
  \bibfield  {author} {\bibinfo {author} {\bibfnamefont {V.}~\bibnamefont
  {Bapst}}\ and\ \bibinfo {author} {\bibfnamefont {G.}~\bibnamefont
  {Semerjian}},\ }\href {\doibase 10.1088/1742-5468/2012/06/P06007} {\bibfield
  {journal} {\bibinfo  {journal} {J. Stat. Mech.}\ }\textbf {\bibinfo {volume}
  {2012}},\ \bibinfo {pages} {P06007} (\bibinfo {year} {2012})}\BibitemShut
  {NoStop}%
\bibitem [{\citenamefont {Caneva}\ \emph {et~al.}(2008)\citenamefont {Caneva},
  \citenamefont {Fazio},\ and\ \citenamefont
  {Santoro}}]{caneva_adiabatic_2008-1}%
  \BibitemOpen
  \bibfield  {author} {\bibinfo {author} {\bibfnamefont {T.}~\bibnamefont
  {Caneva}}, \bibinfo {author} {\bibfnamefont {R.}~\bibnamefont {Fazio}}, \
  and\ \bibinfo {author} {\bibfnamefont {G.~E.}\ \bibnamefont {Santoro}},\
  }\href {\doibase 10.1103/PhysRevB.78.104426} {\bibfield  {journal} {\bibinfo
  {journal} {Phys. Rev. B}\ }\textbf {\bibinfo {volume} {78}},\ \bibinfo
  {pages} {104426} (\bibinfo {year} {2008})}\BibitemShut {NoStop}%
\bibitem [{\citenamefont {Solinas}\ \emph {et~al.}(2008)\citenamefont
  {Solinas}, \citenamefont {Ribeiro},\ and\ \citenamefont
  {Mosseri}}]{solinas_dynamical_2008-1}%
  \BibitemOpen
  \bibfield  {author} {\bibinfo {author} {\bibfnamefont {P.}~\bibnamefont
  {Solinas}}, \bibinfo {author} {\bibfnamefont {P.}~\bibnamefont {Ribeiro}}, \
  and\ \bibinfo {author} {\bibfnamefont {R.}~\bibnamefont {Mosseri}},\ }\href
  {\doibase 10.1103/PhysRevA.78.052329} {\bibfield  {journal} {\bibinfo
  {journal} {Phys. Rev. A}\ }\textbf {\bibinfo {volume} {78}},\ \bibinfo
  {pages} {052329} (\bibinfo {year} {2008})}\BibitemShut {NoStop}%
\bibitem [{\citenamefont {Acevedo}\ \emph {et~al.}(2014)\citenamefont
  {Acevedo}, \citenamefont {Quiroga}, \citenamefont {Rodr\'iguez},\ and\
  \citenamefont {Johnson}}]{acevedo_new_2014}%
  \BibitemOpen
  \bibfield  {author} {\bibinfo {author} {\bibfnamefont {O.~L.}\ \bibnamefont
  {Acevedo}}, \bibinfo {author} {\bibfnamefont {L.}~\bibnamefont {Quiroga}},
  \bibinfo {author} {\bibfnamefont {F.~J.}\ \bibnamefont {Rodr\'iguez}}, \ and\
  \bibinfo {author} {\bibfnamefont {N.~F.}\ \bibnamefont {Johnson}},\ }\href
  {\doibase 10.1103/PhysRevLett.112.030403} {\bibfield  {journal} {\bibinfo
  {journal} {Phys. Rev. Lett.}\ }\textbf {\bibinfo {volume} {112}},\ \bibinfo
  {pages} {030403} (\bibinfo {year} {2014})}\BibitemShut {NoStop}%
\bibitem [{\citenamefont {Hwang}\ \emph {et~al.}(2015)\citenamefont {Hwang},
  \citenamefont {Puebla},\ and\ \citenamefont {Plenio}}]{hwang_quantum_2015-2}%
  \BibitemOpen
  \bibfield  {author} {\bibinfo {author} {\bibfnamefont {M.-J.}\ \bibnamefont
  {Hwang}}, \bibinfo {author} {\bibfnamefont {R.}~\bibnamefont {Puebla}}, \
  and\ \bibinfo {author} {\bibfnamefont {M.~B.}\ \bibnamefont {Plenio}},\
  }\href {\doibase 10.1103/PhysRevLett.115.180404} {\bibfield  {journal}
  {\bibinfo  {journal} {Phys. Rev. Lett.}\ }\textbf {\bibinfo {volume} {115}},\
  \bibinfo {pages} {180404} (\bibinfo {year} {2015})}\BibitemShut {NoStop}%
\bibitem [{\citenamefont {Defenu}\ \emph {et~al.}(2018)\citenamefont {Defenu},
  \citenamefont {Enss}, \citenamefont {Kastner},\ and\ \citenamefont
  {Morigi}}]{defenu_dynamical_2018-2}%
  \BibitemOpen
  \bibfield  {author} {\bibinfo {author} {\bibfnamefont {N.}~\bibnamefont
  {Defenu}}, \bibinfo {author} {\bibfnamefont {T.}~\bibnamefont {Enss}},
  \bibinfo {author} {\bibfnamefont {M.}~\bibnamefont {Kastner}}, \ and\
  \bibinfo {author} {\bibfnamefont {G.}~\bibnamefont {Morigi}},\ }\href
  {\doibase 10.1103/PhysRevLett.121.240403} {\bibfield  {journal} {\bibinfo
  {journal} {Phys. Rev. Lett.}\ }\textbf {\bibinfo {volume} {121}},\ \bibinfo
  {pages} {240403} (\bibinfo {year} {2018})}\BibitemShut {NoStop}%
\bibitem [{\citenamefont {Cirac}\ \emph {et~al.}(1998)\citenamefont {Cirac},
  \citenamefont {Lewenstein}, \citenamefont {M\o{}lmer},\ and\ \citenamefont
  {Zoller}}]{cirac_quantum_1998-1}%
  \BibitemOpen
  \bibfield  {author} {\bibinfo {author} {\bibfnamefont {J.~I.}\ \bibnamefont
  {Cirac}}, \bibinfo {author} {\bibfnamefont {M.}~\bibnamefont {Lewenstein}},
  \bibinfo {author} {\bibfnamefont {K.}~\bibnamefont {M\o{}lmer}}, \ and\
  \bibinfo {author} {\bibfnamefont {P.}~\bibnamefont {Zoller}},\ }\href
  {\doibase 10.1103/PhysRevA.57.1208} {\bibfield  {journal} {\bibinfo
  {journal} {Phys. Rev. A}\ }\textbf {\bibinfo {volume} {57}},\ \bibinfo
  {pages} {1208} (\bibinfo {year} {1998})}\BibitemShut {NoStop}%
\bibitem [{\citenamefont {Pezz\`e}\ \emph {et~al.}(2018)\citenamefont
  {Pezz\`e}, \citenamefont {Smerzi}, \citenamefont {Oberthaler}, \citenamefont
  {Schmied},\ and\ \citenamefont {Treutlein}}]{pezze_quantum_2018-1}%
  \BibitemOpen
  \bibfield  {author} {\bibinfo {author} {\bibfnamefont {L.}~\bibnamefont
  {Pezz\`e}}, \bibinfo {author} {\bibfnamefont {A.}~\bibnamefont {Smerzi}},
  \bibinfo {author} {\bibfnamefont {M.~K.}\ \bibnamefont {Oberthaler}},
  \bibinfo {author} {\bibfnamefont {R.}~\bibnamefont {Schmied}}, \ and\
  \bibinfo {author} {\bibfnamefont {P.}~\bibnamefont {Treutlein}},\ }\href
  {\doibase 10.1103/RevModPhys.90.035005} {\bibfield  {journal} {\bibinfo
  {journal} {Rev. Mod. Phys.}\ }\textbf {\bibinfo {volume} {90}},\ \bibinfo
  {pages} {035005} (\bibinfo {year} {2018})}\BibitemShut {NoStop}%
\bibitem [{\citenamefont {Lucamarini}\ \emph {et~al.}(2004)\citenamefont
  {Lucamarini}, \citenamefont {Paganelli},\ and\ \citenamefont
  {Mancini}}]{lucamariniTwoqubitEntanglementDynamics2004}%
  \BibitemOpen
  \bibfield  {author} {\bibinfo {author} {\bibfnamefont {M.}~\bibnamefont
  {Lucamarini}}, \bibinfo {author} {\bibfnamefont {S.}~\bibnamefont
  {Paganelli}}, \ and\ \bibinfo {author} {\bibfnamefont {S.}~\bibnamefont
  {Mancini}},\ }\href {\doibase 10.1103/PhysRevA.69.062308} {\bibfield
  {journal} {\bibinfo  {journal} {Phys. Rev. A}\ }\textbf {\bibinfo {volume}
  {69}},\ \bibinfo {pages} {062308} (\bibinfo {year} {2004})}\BibitemShut
  {NoStop}%
\bibitem [{\citenamefont {{van Wezel}}\ \emph {et~al.}(2005)\citenamefont {{van
  Wezel}}, \citenamefont {{van den Brink}},\ and\ \citenamefont
  {Zaanen}}]{vanwezelIntrinsicLimitQuantum2005}%
  \BibitemOpen
  \bibfield  {author} {\bibinfo {author} {\bibfnamefont {J.}~\bibnamefont {{van
  Wezel}}}, \bibinfo {author} {\bibfnamefont {J.}~\bibnamefont {{van den
  Brink}}}, \ and\ \bibinfo {author} {\bibfnamefont {J.}~\bibnamefont
  {Zaanen}},\ }\href {\doibase 10.1103/PhysRevLett.94.230401} {\bibfield
  {journal} {\bibinfo  {journal} {Phys. Rev. Lett.}\ }\textbf {\bibinfo
  {volume} {94}},\ \bibinfo {pages} {230401} (\bibinfo {year}
  {2005})}\BibitemShut {NoStop}%
\bibitem [{\citenamefont {Koashi}\ \emph {et~al.}(2000)\citenamefont {Koashi},
  \citenamefont {Bu{\v z}ek},\ and\ \citenamefont
  {Imoto}}]{koashi_entangled_2000}%
  \BibitemOpen
  \bibfield  {author} {\bibinfo {author} {\bibfnamefont {M.}~\bibnamefont
  {Koashi}}, \bibinfo {author} {\bibfnamefont {V.}~\bibnamefont {Bu{\v z}ek}},
  \ and\ \bibinfo {author} {\bibfnamefont {N.}~\bibnamefont {Imoto}},\ }\href
  {\doibase 10.1103/PhysRevA.62.050302} {\bibfield  {journal} {\bibinfo
  {journal} {Phys. Rev. A}\ }\textbf {\bibinfo {volume} {62}},\ \bibinfo
  {pages} {050302} (\bibinfo {year} {2000})}\BibitemShut {NoStop}%
\end{thebibliography}
%

\footnotetext[1]{See Supplemental Material  for  a study of adiabaticity requirements for the ground state preparation, eigenstate spectroscopy at higher energy, a description of the classical mean-field treatment, a derivation of the critical Hamiltonian, and an extensive discussion of parity  symmetry breaking by an external magnetic field.}

\footnotetext[3]{
We found no signature of such a mechanism for the densest clouds achievable in our setup (density $n\simeq \SI{1e14}{cm^{-3}}$). Studying this effect would thus require decreasing the level of magnetic field noise, as the associated Zeeman splitting dominates over the typical dipole-dipole energy scale in our current system.  
}

\footnotetext[4]{
The minimum value $\left< \sigma_{1y} \sigma_{2y}\right>=-0.040(5)$ for $\lambda\simeq\omega_z$ is about 60\% of the minimum correlation $\left<\sigma_{1\hat{\mathbf{n}}} \sigma_{2\hat{\mathbf{n}}}\right>=-1/(N-1)$  allowed for a set of $N$ spins symmetric upon exchange \cite{koashi_entangled_2000}.
}

\footnotetext[5]{
The measured oscillations become randomly dephased for $t\gtrsim\SI{100}{\micro s}$ due to shot-to-shot fluctuations of the laser coupling $\lambda$ and Larmor frequency $\omega_z$, limiting the frequency resolution of our measurements to about $0.05\,\omega_z$.
}

\footnotetext[6]{
It becomes inaccurate in the deep ferromagnetic phase (for $\lambda\gtrsim 1.5\,\omega_z$, see \cite{Note1}), where the precise shape of the double-well potential is required to  describe quantitatively quantum tunneling effects \cite{ulyanov_new_1992}. 
}

\footnotetext[7]{
We present in the Supplementary Material measurements of higher energy levels, that confirm this picture on a larger energy range.
}

\end{document}


\title{Supplementary Material\\
Probing quantum criticality and symmetry breaking at the microscopic level
 }
 \author{Vasiliy Makhalov}
 \thanks{These two authors contributed equally.}
 \author{Tanish Satoor}
 \thanks{These two authors contributed equally.}
 \author{Alexandre Evrard}
 \author{Thomas Chalopin}
 \author{Raphael Lopes}
 \author{Sylvain Nascimbene}
 \email{sylvain.nascimbene@lkb.ens.fr}
 \affiliation{Laboratoire Kastler Brossel,  Coll\`ege de France, CNRS, ENS-PSL University, Sorbonne Universit\'e, 11 Place Marcelin Berthelot, 75005 Paris, France}
 \date{\today}
 
 \maketitle 

\section{Adiabaticity requirements}

\begin{figure}[h!]
\includegraphics[
draft=false,scale=0.9,
trim={5mm 3mm 0 0.cm},
]{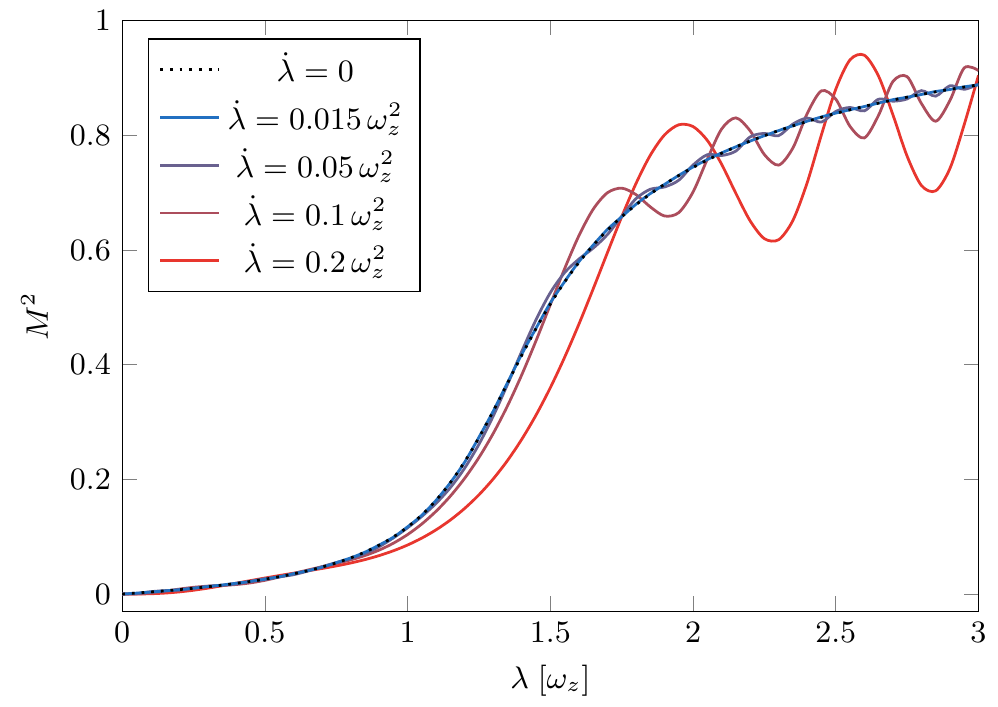}
\caption{
Ferromagnetic correlator $M^2=\langle\sigma_{1x}\sigma_{2x}\rangle$ as a function of $\lambda$ calculated in the state reached after a linear ramp of the coupling  from zero to $\lambda$, for different ramp speeds $\dot\lambda$.
\label{fig_adiabaticity}}
\end{figure}

In order to evaluate the adiabatic character of the ramp used to prepare the ground state, we simulate the system dynamics  by solving the Schr\"odinger equation  numerically with a light coupling increasing at constant speed $\dot\lambda$, as performed, to good approximation, in our experiment. As shown in Fig.\,\ref{fig_adiabaticity} we find that the calculated ferromagnetic correlator $M^2=\langle\sigma_{1x}\sigma_{2x}\rangle$ significantly deviates from the ground state value for ramp speeds $\dot\lambda\gtrsim0.05\,\omega_z^2$. For the ramp speed $\dot\lambda=0.015\,\omega_z^2$ used to prepare the ground state in our experiment, we numerically find that the correlator $M^2$ is practically indistinguishable from the ground-state value, such that the ramp can be considered adiabatic.

\section{Eigenstate spectroscopy}

\begin{figure}
\includegraphics[
draft=false,scale=0.9,
trim={3mm 3mm 0 0.cm},
]{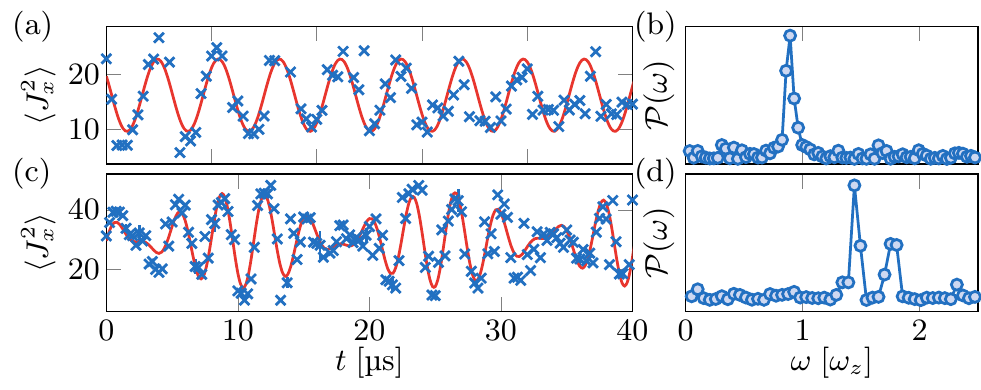}
\caption{
(a,c) Time evolution of the second moment $\langle J_x^2\rangle$ for a coupling $\lambda=\omega_z$, following a weak excitation to populate states $n=0,2$ (a) or a stronger excitation to states $n=2,4,6$ (c). (b,d) Fourier spectra of the (a,c) evolutions.
\label{fig_osc_spec}}
\end{figure} 

We present here a spectroscopy of excited states of the LMGm that goes beyond the measurement of the frequency gaps $\delta$ and $\Delta$ discussed in the main text.
The  spectroscopy is based on the controlled excitation of the system to a state populating several eigenstates, such that the time evolution of the second moment $\langle J_x^2\rangle$ exhibits oscillation frequencies corresponding to the energy spacing between populated levels. 

Starting with all atoms polarized in $\ket{m=-J}_z$ and in the absence of the light beam, we quench the light intensity to induce a coupling $\lexc$ for a duration $\texc$. We then quench to $\lambda=\omega_z$ and let the system evolve for a duration $t$.
We measure the second moment $\langle J_x^2\rangle$ for variable durations $0<t<\SI{100}{\micro s}$ and apply a Fourier transform to the oscillation.
The energy levels $E_n$ are labelled in increasing order by an integer  $0\leq n\leq 2J$, whose parity reflects the quantum state parity.
We relate the measured oscillation frequencies to the spacings $\Delta_{m,n}\equiv (E_n-E_m)/\hbar$.
As the perturbation preserves parity, only even-parity states should be excited.

We show in Fig.\,\ref{fig_osc_spec}a-d two examples of evolutions of the second moment $\langle J_x^2\rangle$ at the critical point $\lambda=\omega_z$, together with their corresponding Fourier spectra.
Fig.\,\ref{fig_osc_spec}a corresponds to a weak excitation, such that the system mostly populates the first two even-parity states, leading to an almost harmonic evolution of frequency $\Delta_{0,2}$. Fig.\,\ref{fig_osc_spec}b corresponds to a stronger excitation with mainly two harmonics whose frequencies are consistent with the spacings $\Delta_{2,4}$ and $\Delta_{4,6}$. 

\begin{figure}
\includegraphics[
draft=false,scale=0.9,trim={5mm 2mm 0 0.cm},
]{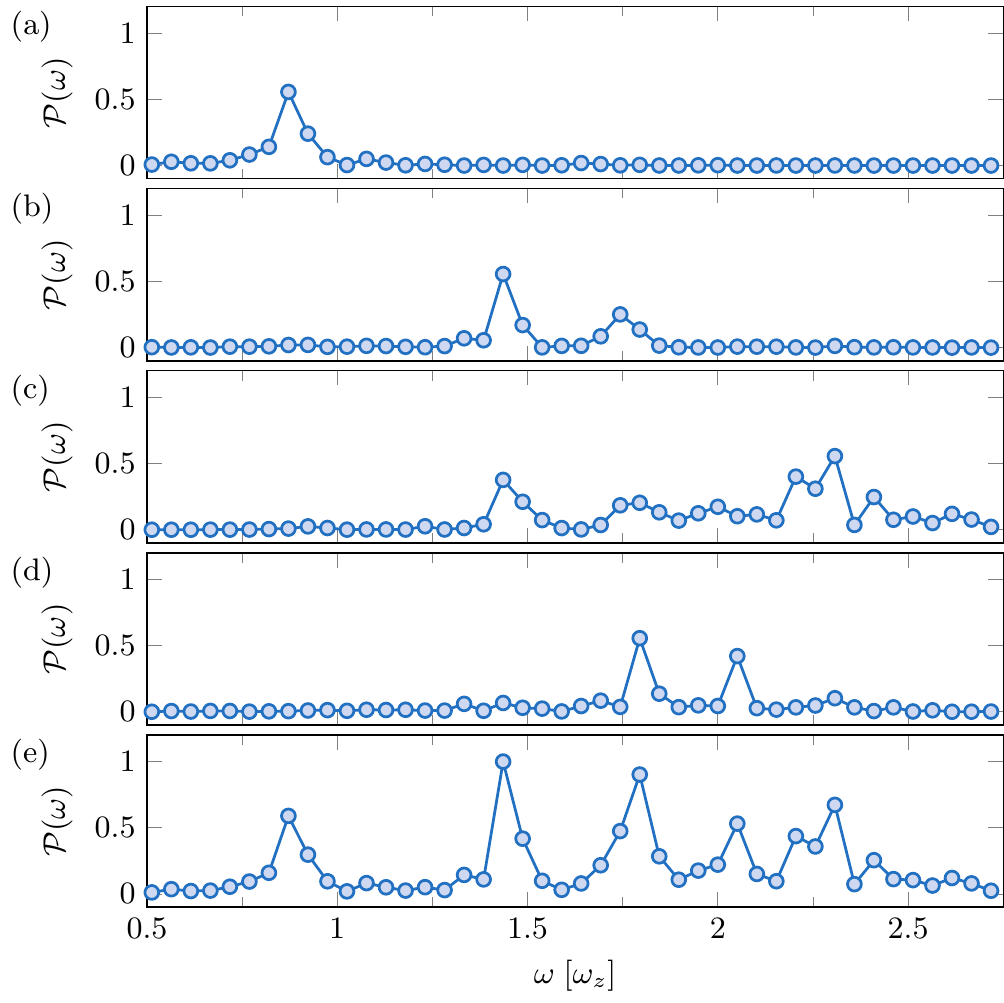}
\caption{
(a-d)  Oscillation spectrum measured at the critical point, after an excitation pulse at $\lexc/\omega_z=$ 1.2 (a), 2.1 (b) 2.66 (c) and 3.25 (d) for a duration $\texc=\SI{11}{\micro s}$ ($\SI{7}{\micro s}$, $\SI{13.5}{\micro s}$ and $\SI{12}{\micro s}$, respectively). (e) Combined spectrum obtained by summing the (a-d) spectra.
\label{fig_spectroscopy}}
\end{figure}

We decrease the background noise level by applying a Gaussian filter of r.m.s. width $0.5\,\omega_z$, centered on the most intense Fourier frequency. We show in Fig.\,\ref{fig_spectroscopy}a-d four Fourier spectra corresponding to different excitations, which are combined to a final spectrum shown in Fig.\,\ref{fig_spectroscopy}e.

\begin{figure}
\includegraphics[
draft=false,scale=0.9,
trim={3mm 6mm 0 0.cm},
]{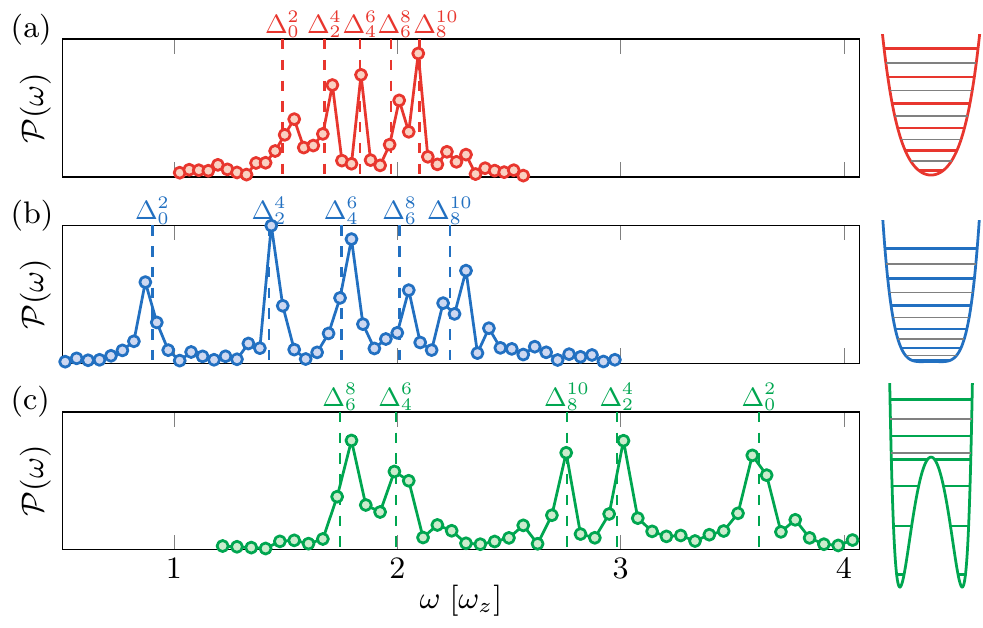}
\vspace{-0mm}
\caption{
(a-c) Excitation Fourier spectra measured for $\lambda/\omega_z=0.50(2)$ (a, red dots), 1.00(2) (b, blue dots) and 3.8(1) (c, green dots). The vertical dashed lines correspond to the frequency spacings $\Delta_{n,n+2}$ calculated from the LMGm. The right panels show the effective potentials and energy spectra for each interaction strength. The thin gray lines indicate odd-parity states, not excited with this protocol.
\label{fig_specs}}
\end{figure}

\begin{figure}
\includegraphics[
draft=false,scale=0.9,
trim={3mm 6mm 0 0.cm},
]{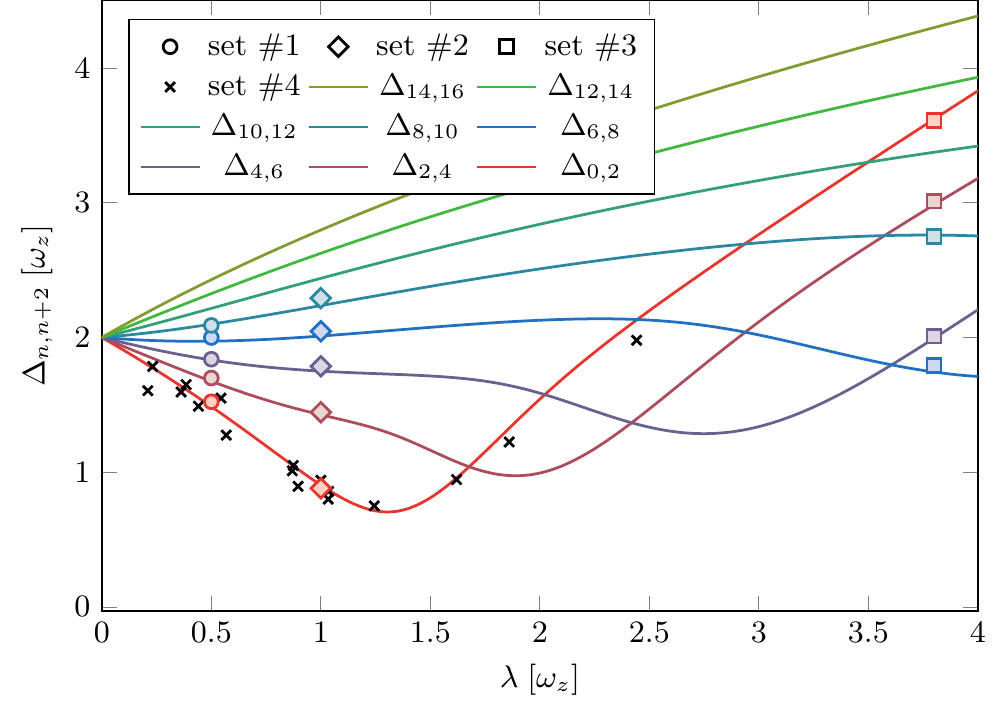}
\vspace{-0mm}
\caption{
Measured frequency spacings  (markers) compared to the LMGm (solid lines). Data sets $\#1$, $\#2$, $\#3$ correspond to interaction strengths $\lambda/\omega_z=0.50(2)$ (circles), 1.00(2) (diamonds) and 3.8(1) (squares), respectively.  Set $\#4$ (crosses) correspond to the frequency gap $\Delta$ shown in Fig.\,3a of the main text. 
\label{fig_spacings}}
\end{figure}

Such spectra were measured for various interaction strengths $\lambda/\omega_z=0.5,1$ and $3.8$ (see Figs.\,\ref{fig_specs}a,b,c) and we identify the spacings  $\Delta_{n,n+2}$ ($n\leq 8$) by comparison with the LMGm (see Fig.\,\ref{fig_spacings}). Hereafter, we interpret these spectra in terms of particle motion in effective potentials  \cite{ulyanov_new_1992}. In the paramagnetic phase ($\lambda=0.5\,\omega_z$), the measured spacings remain close to each other, corresponding to an almost harmonic trap \cite{holsteinFieldDependenceIntrinsic1940,dusuel_finite-size_2004} (see Fig.\,\ref{fig_specs}a). At the critical point $\lambda=\omega_z$, the successive spacings $\Delta_{n,n+2}$ increase with $n$ more significantly (see Fig.\,\ref{fig_specs}b), as expected for a particle evolving in the purely quartic potential (see the critical Hamiltonian (3) in the main text with $\epsilon=0$) \cite{hioeQuantumTheoryAnharmonic1978,leyvraz_large-$n$_2005}. This non-harmonic behavior illustrates the generic property of quantum critical systems, whose low-energy spectra cannot be simply reduced to non-interacting excitation quanta  \cite{sachdev2011quantum}.  Deep in the ferromagnetic phase ($\lambda=3.8\,\omega_z$, Fig.\,\ref{fig_specs}c), the spacings are not ordered monotonically and exhibit a minimum between the 6\textsuperscript{th} and 8\textsuperscript{th} levels (see also Fig.\ref{fig_spacings}). To explain this behavior, we notice that these energy levels are close to the tip of the corresponding double-well potential (see right panel of Fig.\,\ref{fig_specs}c). Our observations are reminiscent of the divergence of the density of states at the tip of a macroscopic double-well potential \cite{leyvraz_large-$n$_2005}, expressing  the occurrence of an excited-state quantum phase transition in the TL \cite{leyvraz_large-$n$_2005,heylDynamicalQuantumPhase2018}.

\section{Symmetry breaking and magnetic susceptibility}

In the thermodynamic limit, the LMGm exhibits a ferromagnetic phase at $\lambda  > \omega_z$, with non-zero order-parameter and a broken $\mathbb{Z}_2$ symmetry, due to the presence of two degenerate ground state of opposite parity $P_z$. For finite size systems, the degeneracy is lifted and the transition is smoothened. The presence of a magnetic field along $x$ with an amplitude comparable to the energy difference between the two lowest states of opposite parity $\hbar\delta$ is then required to break the symmetry.

\begin{figure}[t!]
\includegraphics[
draft=false,scale=0.9,trim={5mm 2mm 0 0.cm},
]{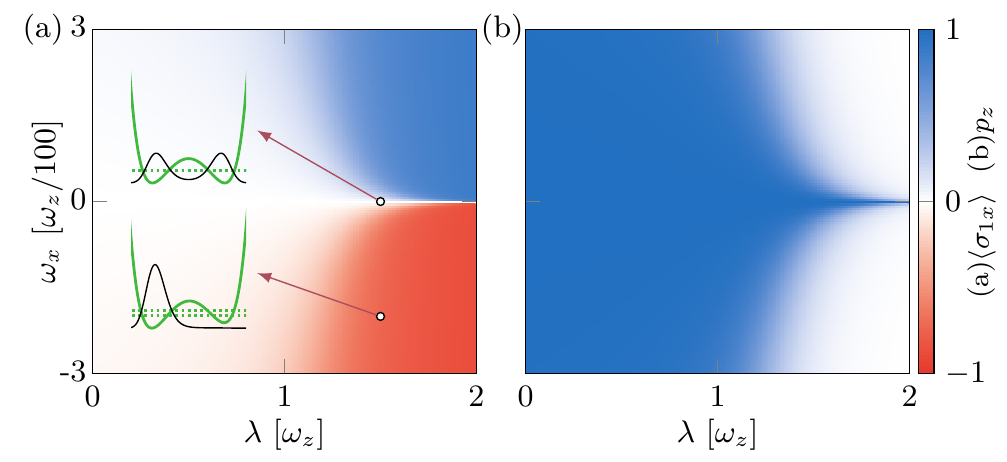}
\caption{
The expectation value of the (a) order parameter $\langle \sigma_{1x}\rangle$ and (b) the parity $p_z$ of the ground state of the $J=8$ LMGm as a function of the ferromagnetic coupling $\lambda$ and the transverse Zeeman coupling $\omega_x$. The insets show the equivalent single particle potentials (solid green), the energies of the ground and first excited states (dashed green), and the squared modulus of the ground state wavefunction (black) for the points $(\lambda / \omega_z, \omega_x/ \omega_z) = (1.5, 0)$ and $(1.5, -0.02)$ respectively. 
\label{fig_SBtheory}}
\end{figure}

\begin{figure}
\includegraphics[
draft=false,scale=0.9,trim={5mm 2mm 0 0.cm},
]{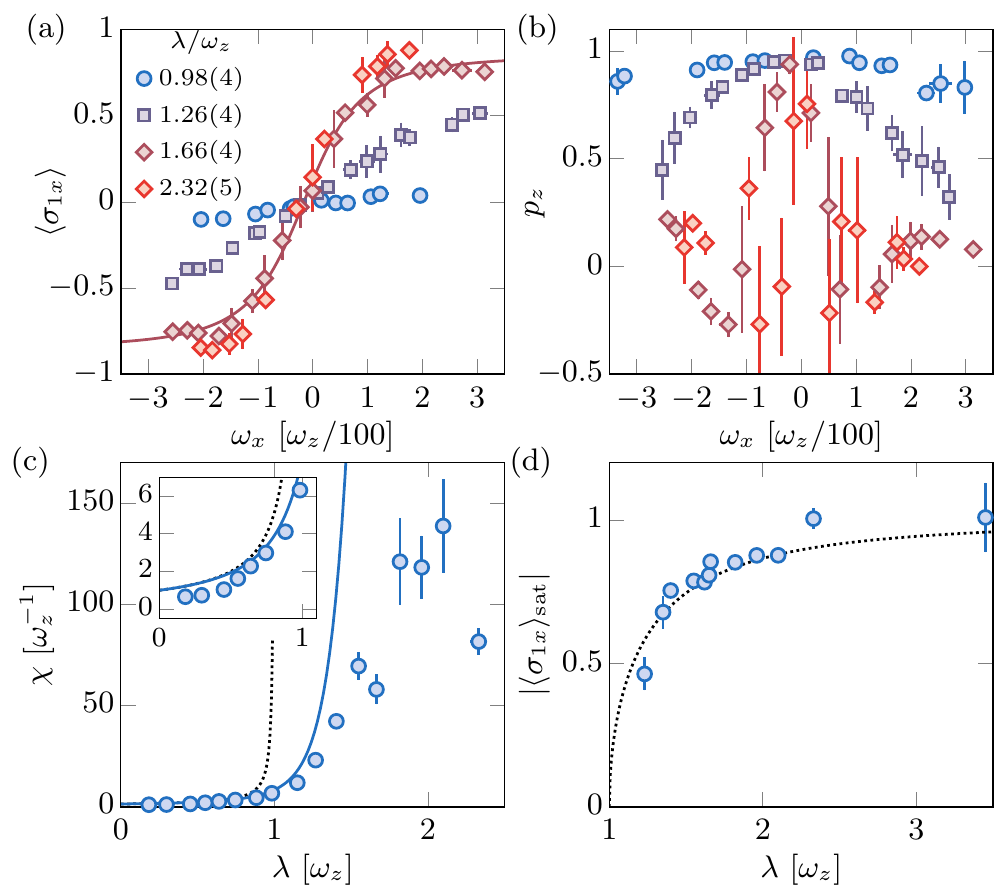}
\caption{
(a) Order parameter $\langle\sigma_{1x}\rangle$ and (b) mean parity $p_z$  as a function of the transverse magnetic field $\omega_x$ for different couplings $\lambda$. The solid line is an example (for $\lambda = 1.66(4)$) of a fit used to extract the saturation values presented in (d). (c) The susceptibility $\chi$ compared to the LMGm (solid blue) and the mean-field values (dotted black). The inset is a zoom-in on small $\chi$s. (d) The fitted value the order parameter saturates to, $\langle\sigma_{1x}\rangle_{\text{sat}}$, when a field is applied along $x$, compared to the mean-field order parameter at zero field (dotted black).
\label{fig_symBreaking}}
\end{figure}

The effective two-well potential provides a visual understanding. The presence of a transverse field along $x$, with amplitude  $\omega_x \gtrsim \delta$, tilts the double well and mixes both states leading to a parity loss (see Fig.\,\ref{fig_SBtheory}a, inset). The wavefunction simultaneously localises in the lower well, and the order parameter acquires a non-zero value.
This is consistent with the results shown in Fig.\,\ref{fig_symBreaking}(a, b) where the correlation between the appearance of a non-zero order-parameter and a parity loss, as a function of $\omega_x$, is revealed.  The small negative parity values shown in Fig.\,\ref{fig_symBreaking}b are consistent with our finite ramp speed $\dot{\lambda} = 0.015 \, \omega_z^2$. 

Experimentally, the populations are extracted from a projective Stern-Gerlach experiment in the presence of a bias field pointing along a direction $\mathbf{\hat{u}} = \px{\omega_x,0,\omega_z}/ {\sqrt{\omega_x^2 + \omega_z^2}}$. For $|\omega_x|\ll\omega_z$, the order-parameter and susceptibility along $\mathbf{\hat u}$ are given by $\langle\sigma_{1x}\rangle^{\hat{{\mathbf{ u}}}} = \langle\sigma_{1x}\rangle - \omega_x/\omega_z \langle\sigma_{1z}\rangle$ and $\chi^{\hat{{\mathbf{ u}}}} = \chi - \langle\sigma_{1z}\rangle / \omega_z$ respectively. The value of $\av{\sigma_{1z}}$ is independently measured and both order-parameter and susceptibility recovered as a function of $\omega_x$ (see Fig.\ref{fig_symBreaking}(a, c)).

In this two-level effective model,  the order parameter obeys the relation
\begin{equation}\label{eq_fittingFuncSat}
\langle\sigma_{1x}\rangle = \frac{\omega_x \av{\sigma_{1x}}_{\text{sat}}}{\sqrt{\omega_x^2+a^2}},
\end{equation}
where $a$ is a free parameter proportional to $\hbar\delta$ and $\av{\sigma_{1x}}_{\text{sat}}$ the saturated value of the order parameter. This ansatz matches our data well for $\lambda>1.2\,\omega_z$ (see Fig.\,\ref{fig_symBreaking}a), and allows us to recover the dependence of $\langle\sigma_{1x}\rangle_{\text{sat}}$ on $\lambda$, which is in good agreement with the mean-field prediction (see Fig.\,\ref{fig_symBreaking}d). This two-level approximation becomes inaccurate for $\lambda\lesssim1.2\,\omega_z$, since the parity gap becomes comparable to the energy scale of higher energy levels.

The susceptibility  $\chi = \text{d} \av{\sigma_{1\, x}}/\text{d} \omega_x |_{\omega_x=0}$ quantifies the response of the system to the presence of a vanishing magnetic field, and diverges at the phase transition in the TL.
We show in Fig.\,\ref{fig_symBreaking}c the susceptibility $\chi$ as a function of $\lambda$. Although our experimental results are in good agreement with the LMGm in the paramagnetic phase we observe an enhanced deviation as a function of $\lambda$. This is attributed to the presence of magnetic field fluctuations, which for $\lambda > 1.5\, \omega_z$ are comparable to the parity gap $\hbar\delta$.

\section{Classical mean-field theory}

The Hamiltonian describing the LMGm can be written as
\begin{equation}
H=\frac{\hbar}{2}\sum_{i=1}^N\left[-\lambda\left(\frac{1}{N-1}\sum_{j\neq i} \frac{\sigma_{jx}}{2}\right)\sigma_{ix}+\omega_z \sigma_{iz}\right].
\end{equation}
The spin $i$ is thus subjected to an external $z$ field and to an $x$ field proportional to the mean spin projection along $x$ of all other spins. The LMGm can thus be viewed as a `quantum mean field' model.

The classical mean-field approximation consists in neglecting the fluctuations of the mean field acting on each spin, via the replacement 
\begin{align}
\sigma_{ix}\sigma_{jx}&\rightarrow  \langle \sigma_{jx} \rangle \sigma_{ix}+ \langle \sigma_{ix} \rangle\sigma_{jx}-\langle \sigma_{ix} \rangle\langle \sigma_{jx} \rangle,\\
&=\langle \sigma_{1x} \rangle \sigma_{ix}+ \langle \sigma_{1x} \rangle\sigma_{jx}-\langle \sigma_{1x} \rangle^2,
\end{align}
where we use the relation  $\langle \sigma_{iu} \rangle=\langle \sigma_{1u} \rangle$ based on the exchange symmetry of the LMGm. We obtain the mean-field Hamiltonian (up to an overall energy shift)
\begin{equation}
H=\frac{\hbar}{2}\sum_{i=1}^N\left(-\lambda \langle \sigma_{1x} \rangle \sigma_{ix}+\omega_z \sigma_{iz}\right),
\end{equation}
corresponding to a system of independent spins in an external magnetic field $\mathbf{B}\propto -\lambda\langle \sigma_{1x}\rangle\unitx+\omega_z\unitz$. The ground state thus corresponds to a paramagnetic state anti-aligned with this field, of projection along $x$
\begin{equation}
\langle \sigma_{1x}\rangle=\frac{\lambda \langle \sigma_{1x} \rangle}{\sqrt{(\lambda \langle \sigma_{1x} \rangle)^2+\omega_z^2}},
\end{equation}
allowing to infer the $x$ spin projection self-consistently, as
\begin{align}
 \langle \sigma_{1x} \rangle &=0&\;\text{ for }\;\lambda<\omega_z, \\
 &= \pm\sqrt{1-\left(\frac{\omega_z}{\lambda}\right)^2}&\;\text{ for }\;\lambda\geq\omega_z.
\end{align}

We also give for completeness the expression of the parity gap $\hbar\delta$ and dynamical gap $\hbar\Delta$ within classical mean-field theory \cite{botet_size_1982,dusuel_continuous_2005-1}, as
\begin{align}
 \frac{\delta}{\omega_z}&=\sqrt{1-\frac{\lambda}{\omega_z}}\;&\text{ for }\;\lambda<\omega_z, \\
&=0\;&\text{ for }\;\lambda\geq\omega_z,
\end{align}
and
\begin{align}
 \frac{\Delta}{\omega_z}&=2\sqrt{1-\frac{\lambda}{\omega_z}}\;&\text{ for }\;\lambda<\omega_z, \\
&=2\sqrt{\left(\frac{\lambda}{\omega_z}\right)^2-1}\;&\text{ for }\;\lambda\geq\omega_z.
\end{align}
The expressions of  the mean parity $p_z$ of the mean-field ground state are
\begin{align}
p_z &=1&\;\text{ for }\;\lambda<\omega_z, \\
&  = (\omega_z/\lambda)^{2J} &\;\text{ for }\;\lambda\geq\omega_z,
\end{align}
which, in the limit $J\gg1$, falls rapidly to zero in the ferromagnetic phase.

\section{Critical low-energy Hamiltonian}

We use a Holstein-Primakoff transformation to express the spin operators in terms of bosonic creation and annihilation operators $a$ and $a^\dagger$, as \cite{holsteinFieldDependenceIntrinsic1940,ulyanov_new_1992}
\begin{align}
    	J_z&=-J+a^\dagger a,
\label{eq_spin}\\
    	J_+&=\sqrt{2J}a^\dagger\sqrt{1-a^\dagger a/(2J)},
\label{eq_spin}\\
    	J_-&=\sqrt{2J}\sqrt{1-a^\dagger a/(2J)}a.
\label{eq_spin2}
\end{align}
In the critical regime, the system remains close to the classical paramagnetic ground state $\ket{m=-J}_z$, such that the number $n$ of bosonic excitations remains much smaller than $J$. We perform a Taylor expansion of (\ref{eq_spin}) and (\ref{eq_spin2}) in powers of $1/J$, that allows us to rewrite the LMG Hamiltonian as
\begin{align}
    H=&-J-1/2\label{eq:constant}\\
    &+\frac{1}{J^{1/3}}\left(\frac{1}{2}P^2+\frac{1}{8}X^4-\frac{\epsilon}{2}X^2\right)
    \label{eq:Hc}\\ 
    &-\frac{1}{J^{2/3}}\left(\frac{1}{4}X^2\right)\label{eq_a1}\\
    &+\frac{1}{J}\left(\frac{1}{32}\{X,\{X,P^2\}\}+\frac{\epsilon}{8}X^4\right)\label{eq_a2}\\
    &-\frac{1}{J^{4/3}}\frac{\epsilon X^2}{4}\label{eq_a3}\\
    &+\mathcal{O}(1/J^{5/3}),
\end{align}
where we define $\epsilon=J^{2/3}(\lambda/\omega_z-1)$ and
 the effective position, $X$, and momentum, $P$, operators as
\begin{align}
    a&=(J^{1/6}X+\I J^{-1/6}P)/\sqrt{2},\\
	a^\dagger &=(J^{1/6}X-\I J^{-1/6}P)/\sqrt{2},
\end{align}
such that $[X,P]=\I$.
Besides the constant $(-J-1/2)$, the low-energy Hamiltonian is dominated in the limit $J\gg1$ by the critical Hamiltonian 
\begin{equation}
H_{\mathrm{c}}=\frac{1}{2}P^2+\frac{1}{8}X^4-\frac{\epsilon}{2}X^2,
\end{equation}
which describes the motion of a massive particle in a quadratic plus quartic potential.

To estimate the regime of applicability of the critical Hamiltonian, we calculate the mean number $n$ of excitations and compare it to $J$. In the paramagnetic phase $\epsilon<0$, the potential has a single minimum for $X=0$, leading to a small amount of excitations in the ground state. Conversely, in the ferromagnetic phase $\epsilon>0$, the potential exhibits two minima for $\pm X_0$, where $X_0=\sqrt{2\epsilon}$, leading to $n\simeq J^{1/3}X_0^2$ excitations in the ground state. We expect this description to be valid as long as $n\lesssim J/2$, i.e. $\epsilon\lesssim J^{2/3}/4$ or, equivalently
\begin{equation}\label{eq_1}
\lambda/\omega_z-1\lesssim 1/4.
\end{equation}
This condition can be recovered more rigorously by direct evaluation of  the next-order terms (\ref{eq_a1}), (\ref{eq_a2}) and (\ref{eq_a3}).

\begin{figure}
\includegraphics[
draft=false,scale=1.0,trim={5mm 2mm 0 0.cm},
]{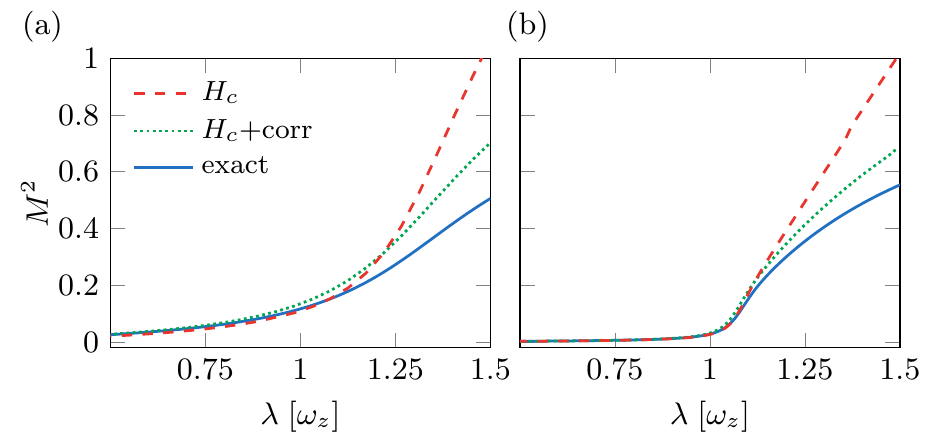}
\caption{Evolution of the ferromagnetic correlator $M^2$ with $\lambda$ for the LMGm (blue line), the critical Hamiltonian (dashed red line), the critical Hamiltonian with higher-order corrections (green dotted line), for
(a) $J=8$ and (b) $J=80$.
\label{Fig:Mx2:j8:j80}}
\end{figure}

We confirm this behavior by calculating the ferromagnetic correlator $M^2$ in the ground state of the LMGm, critical Hamiltonian and critical Hamiltonian plus additional expansion terms, for $J=8$ and $J=80$. As shown in Fig.\,\ref{Fig:Mx2:j8:j80}, the critical Hamiltonian deviates from the LMGm for $\lambda\simeq1.2\,\omega_z$, in agreement with (\ref{eq_1}).

\include{SupplementaryMaterial.bbl}